\documentclass[acmsmall]{acmart} 
\newcommand{\xingbo}[1]{{\color{black}{#1}}}
\newcommand{\finalchange}[1]{{\color{black}{#1}}}
\newcommand{\sentence}{1,269 }

\newcommand{\nonarguments}{102 }
\newcommand{\premise}{1,034 }
\newcommand{\claim}{235 }
\newcommand{\logos}{583 }
\newcommand{\pathos}{135 }
\newcommand{\evidence}{525 }
\newcommand{\ethos}{79 }

\usepackage{multirow}
\usepackage{booktabs}
\usepackage{array}
\usepackage{graphicx}

\AtBeginDocument{%
  \providecommand\BibTeX{{%
    \normalfont B\kern-0.5em{\scshape i\kern-0.25em b}\kern-0.8em\TeX}}}

\setcopyright{acmcopyright}
\copyrightyear{2022}
\acmYear{2022}

\acmJournal{PACMHCI}

\begin{document}

\title{Persua: A Visual Interactive System to Enhance the Persuasiveness of Arguments in Online Discussion}


\author{Meng Xia}
\affiliation{%
  \institution{Carnegie Mellon University}
  \city{Pittsburgh}
  \country{United States}}
\email{iris.xia@connect.ust.hk}

\author{Qian Zhu}
\affiliation{%
  \institution{The Hong Kong University of Science and Technology}
  \city{Hong Kong}
  \country{China}}
\email{qian.zhu@connect.ust.hk}

\author{Xingbo Wang}
\affiliation{%
  \institution{The Hong Kong University of Science and Technology}
  \city{Hong Kong}
  \country{China}}
\email{xingbo.wang@connect.ust.hk}

\author{Fei Nie}
\affiliation{%
  \institution{The Hong Kong University of Science and Technology}
  \city{Hong Kong}
  \country{China}}
\email{fnie9421@gmail.com}

\author{Huamin Qu}
\affiliation{%
  \institution{The Hong Kong University of Science and Technology}
  \city{Hong Kong}
  \country{China}}
\email{huamin@cse.ust.hk}

\author{Xiaojuan Ma}
\affiliation{%
  \institution{The Hong Kong University of Science and Technology}
  \city{Hong Kong}
  \country{China}}
\email{mxj@cse.ust.hk}

\renewcommand{\shortauthors}{Trovato and Tobin, et al.}

\begin{abstract}
Persuading people to change their opinions is a common practice in online discussion forums on topics ranging from political campaigns to relationship consultation. Enhancing people's ability to write persuasive arguments could not only practice their critical thinking and reasoning but also contribute to the effectiveness and civility in online communication. It is, however, not an easy task in online discussion settings where written words are the primary communication channel. In this paper, we derived four design goals for a tool that helps users improve the persuasiveness of arguments in online discussions through a survey with 123 online forum users and interviews with five debating experts. To satisfy these design goals, we analyzed and built a labeled dataset of fine-grained persuasive strategies (i.e., \textit{logos}, \textit{pathos}, \textit{ethos}, and \textit{evidence}) in 164 arguments with high ratings on persuasiveness from ChangeMyView, a popular online discussion forum. We then designed an interactive visual system, Persua, which provides example-based guidance on persuasive strategies to enhance the persuasiveness of arguments. In particular, the system constructs portfolios of arguments based on different persuasive strategies applied to a given discussion topic. It then presents concrete examples based on the difference between the portfolios of user input and high-quality arguments in the dataset. A between-subjects study shows suggestive evidence that Persua encourages users to submit more times for feedback and helps users improve more on the persuasiveness of their arguments than a baseline system. Finally, a set of design considerations was summarized to guide future intelligent systems that improve the persuasiveness in text.
\end{abstract}

\ccsdesc[500]{Applied computing~Interactive learning environments}
\ccsdesc[500]{Computing methodologies~Natural language processing}
\ccsdesc[500]{Human-centered computing~Interactive systems and tools}

\keywords{persuasive strategies, example-based learning, argumentation learning, educational applications.}

\maketitle
\section{Introduction}
It is very common that people hold different stances on a topic and try to persuade others to change their opinions during social interactions online~\cite{tan2016winning}. We can see a massive scale of interpersonal persuasion on various social media platforms over topics ranging from political campaigns, business pitches to daily issues, including but not limited to relationship consultation~\cite{Fogg2008Persuasion}. There even emerge dedicated online forums such as Debate\footnote{\url{https://www.debate.org/}} and ChangeMyView\footnote{\url{https://www.reddit.com/r/changemyview/}\label{CMV}} for people to communicate and induce others to believe their opinions through written words~\cite{CMV_2019NPR}. While many people are actively practicing interpersonal persuasion in online discussion forums (\emph{e.g.}, ChangeMyView\textsuperscript{\ref{CMV}} has over 700,000 subscribers), it is not an easy task to write truly effective persuasive arguments to change other people's views~\cite{durmus2019exploring}. Enhancing people's ability to write persuasive arguments could not only practice their critical thinking and reasoning~\cite{AL-CHI2020}, but also contribute to the effectiveness and civility in online communication~\cite{goovaerts2020uncivil}. However, it is still an open question as to the best way to help people enhance their persuasiveness of arguments in online discussion.




There has been an extensive body of literature on persuasion theories~\cite{cialdini2007influence, dillard2002persuasion, eagly1993psychology, popkin2020reasoning, reardon1991persuasion}, many of which target verbal, face-to-face communication contexts~\cite{chatterjee_verbal_2014, Talk_likeTED14}.
It is hence difficult for people to apply persuasive theories in an online discussion setting where written words are the primary, if not the only, communication channel~\cite{connors1979differences}.
Some earlier works on persuasive writing focus on analysis about persuasive arguments in business or charity-related scenarios such as fund-raising campaigns~\cite{Zhou15, CHI-EA13_crowdfundingTool, p2p_business, xiao2019should, yang2017persuading}.
They are not practical enough to guide people to improve their interpersonal persuasion on casual topics in online forums~\cite{tan2016winning}.
Recently, an increasing number of methods and tools have been proposed to help people refine written arguments with natural language processing (NLP) techniques in a broader range of contexts~\cite{afrin2021effective}. For example, AL, introduced by human-computer interaction researchers, is an adaptive learning tool that provides feedback on the argumentation structure to students' writing~\cite{AL-CHI2020}. 
These existing tools focus mainly on analyzing the structure of arguments (i.e., the relationship of claims and premises) while lacking guidance on persuasive strategies (e.g., logistic reasoning, emotional affection)~\cite{carlile-etal-2018-give} that can be applied to written arguments.
\xingbo{Previous literature~\cite{gallo2014talk, han2019appeals, ting2018ethos, verma2020pathos} has provided empirical evidence that the composition of different exploited persuasive strategies may affect the persuasiveness of written arguments. For example, when analyzing the informal requests of 99 university students, emotional appeal (\textit{pathos}) were used the most and were more successful than rational appeal (\textit{logos}) or credibility appeal (\textit{ethos}) ~\cite{ting2018ethos}.}
In addition, these works have not made full use of enormous discussion data collaboratively generated by peers in the online discussion environment. Showing persuasive strategies and successful arguments from peers might encourage people to reflect on their writing and construct more convincing arguments.

\xingbo{To fill this gap, we conducted a need-finding survey with 123 online forum users to understand their demands and challenges on improving arguments persuasiveness in the online discussion. As a triangulation, we conducted semi-structured interviews with five domain experts who have rich experience in debate competitions and knowledge on persuasive theories for suggestions that can address users' needs. Overall, 121 out of 123 users expressed a desire for a tool to help them improve the persuasiveness in written texts.
Most of these participants find it hard to come up with appropriate examples, adapt the language style (e.g., professional language style or emotionally appealing language style) to the topic, and improve the logic of arguments.
By discussing these findings with experts, we mapped language styles from users' responses to different compositions of persuasive strategies (i.e., \textit{logos}, \textit{pathos}) in persuasive theories (e.g., Aristotle's theory of persuasion~\cite{rapp2011aristotle}) and built a taxonomy of persuasive arguments, which comprises two types of argument components (i.e., claim and premise) and four different persuasive strategies (i.e., \textit{pathos}, \textit{logos}, \textit{ethos}, and \textit{evidence}). In addition, by further referring to learning theories of example-learning~\cite{schworm2007learning, van2010example, xia2019peerlens,  leavitt2017upvote, wang2020voicecoach} and cognitive dissonance theory~\cite{festinger1962cognitive} (i.e., people would become uncomfortable and introspective when their existing knowledge conflicts with the information presented), we derived the major design goals of the tool to improve writing persuasiveness: (1) providing examples of high-quality persuasive arguments fitting to the topic and support filtering by persuasive strategies; (2) displaying the composition of different persuasive strategies in the user's input and compare it with other well-accepted persuasive arguments; (3) showing the logical structure of the arguments; and (4) offering a web-based interface and visual augmented feedback for the arguments writing.}

To satisfy the design goals, we first analyzed and built a labeled dataset of fine-grained persuasive strategies in 164 arguments (\sentence sentences) with high ratings on persuasiveness from ChangeMyView\textsuperscript{\ref{CMV}}, a subreddit dedicated to civil discourse. We then trained a series of machine learning models on this labeled dataset to automatically detect persuasive strategies in written documents. Next, we developed an interactive visual system called Persua to provide example-based guidance on persuasive strategies to help users strengthen their arguments. In particular, the system constructs portfolios of successful arguments (as rated in ChangeMyView\textsuperscript{\ref{CMV}}) to characterize the composition of different persuasive strategies in response to a given discussion topic. It then plots these persuasive portfolios onto a 2D space and visualizes the difference in the portfolio of their own arguments for users to compare. Based on the comparison, the system provides concrete examples of different persuasive strategies for users' reference. Finally, we conducted a between-subjects study with 36 university students, inexperienced at writing persuasive arguments online, to compare Persua with a baseline in terms of efficacy on improving arguments' persuasiveness, intuitiveness of visual design, and system usability. Suggestive evidence shows that Persua encourages participants to submit arguments more times for feedback and improve the persuasiveness of the arguments more than Baseline. Participants also rated Persua significantly more informative than Baseline.
We summarized our main contributions as follows:
\begin{itemize}
\item
An interactive visual system that assists people to enhance the persuasiveness in their arguments in the online discussion. In particular, it constructs portfolios of fine-grained persuasive strategies in arguments and provides concrete examples of different persuasive strategies based on discussion data collaboratively generated.
\item
A between-subjects study that demonstrates the effectiveness and usefulness of Persua compared with a baseline system.
\item
A set of design considerations that guide the future design of the interactive system assisting people in improving the persuasiveness of written texts for online discussion.
\end{itemize}
\section{Related Work}
This section introduces the literature on persuasive strategies in general, persuasion analysis in the online discussion, and computer-aided argumentation learning systems.

\xingbo{\subsection{Persuasive Strategies in General}
Persuasive strategies are devices people apply in communication to persuade others~\cite{schmitz2012primer}, which date back to Aristotle. Aristotle's Rhetoric is an ancient Greek treatise on the art of persuasion, developed in the 4th-century BC~\cite{rapp2011aristotle}. In the treatise, Aristotle proposed three persuasive strategies: \textit{ethos}, \textit{pathos}, and \textit{logos}. \textit{Ethos} means arguments that build the speaker's credibility by sharing professional experience; \textit{pathos} indicates persuading people by arousing their emotions; and \textit{logos} represents arguments that use facts and logical reasoning. This theory has been widely adopted in diverse domains beyond western countries~\cite{adhikary2010explorations}. Based on this theory, Carmine Gallo developed nine practical skills to make public speaking persuasive, including ``Favor pictures over text'' and ``Sticking to the 18-minute Rule''~\cite{gallo2014talk}. Robert B. Cialdini introduced six principles on how to enhance persuasion and influence~\cite{cialdini2007influence}. For example, he added ``Social Proof'' (i.e., ``\textit{when we are unsure, we look to similar others to provide us with the correct actions to take. The more people who undertake that action, the more we consider that action correct.}'').
Kathleen Reardon~\cite{reardon1991persuasion} demonstrates applications of persuasion theory, incorporating Aristotle's theory, in four communication contexts--interpersonal, organizational, mass media, and political. In addition, many researchers investigated the application of Aristotle's theories in other domains, including public speaking~\cite{dlugan2010ethos, Talk_likeTED14}, courtroom discussions~\cite{mccormack2014ethos}, and online hotel booking platforms~\cite{han2019appeals}. These studies showed the effectiveness of Aristotle's theory or its variations.

\finalchange{In this work, we utilize existing theories, especially Aristotle's Rhetoric, as fundamental building blocks to investigate how to employ persuasive strategies properly in writing scenarios, especially online discussions.}}


\subsection{Persuasion Analysis in Online Discussion}
Previous works on persuasion in online discussion mainly focus on which factors affect the prediction outcome---whether people are persuaded or not~\cite{hample2006toward}. These factors can be classified into three categories: argument information~\cite{boltuvzic2014back, basave2016study, habernal2016makes, gleize2019you, wang2019persuasion, yang2019let, egawa2019annotating}, social interactions information~\cite{tan2016winning, durmus2019modeling}, and audience background
information~\cite{longpre2019persuasion, durmus2019exploring}. Among them, argument content analysis attracts more attention, where most studies are conducted. 

\xingbo{As for arguments analysis, previous work has mainly investigated the structure of the written argument~\cite{lawrence2020argument}. Toulmin model~\cite{brockriede1960toulmin}, developed by philosopher Stephen E. Toulmin, is a famous taxonomy of argumentation structure, in which arguments are categorized into six parts: claim, grounds, warrant, qualifier, rebuttal, and backing.} Researchers progressively detected more structural components based on Toulmin model in the online discussion context. Filip and Jan first proposed the task, argument recognition, to detect arguments and non-arguments in an online discussion as they are vague and poorly worded~\cite{boltuvzic2014back}. Amparo Elizabeth Cano-Basave and Yulan further classified the arguments into claims and premises, and then analyzed how the support or attack relationship between claim and premise influences the persuasive effect in the context of political debates~\cite{basave2016study}. To improve the prediction accuracy and understand why one argument is more persuasive than another, Ivan Habernal and Iryna Gurevych conducted an empirical analysis to ask people to write down the reasons by comparing each pair of arguments~\cite{habernal2016makes}. They summarized 26000 unique reasons in nine classes. Following this work, the IBM research team took a step into focusing on those reasons and evidences instead of the claim, using a Siamese neural network to achieve higher accuracy in predicting which argument is more convincing~\cite{gleize2019you}.

With the development of NLP techniques, more recent work goes beyond the structure of the argument and investigates more detailed persuasive strategies from the content perspective in the online dialogue. For example, analyzing the persuasive strategies when persuading people to donate money to a charity~\cite{wang2019persuasion, yang2019let}. 
However, they are not practical enough to guide people to improve their persuasiveness in an online discussion setting over diverse topics.

\subsection{Computer-aided Argumentation Learning}
Relevant research on improving the persuasiveness in online discussion is on computer-aided argumentation learning. These systems and interfaces can be divided into three aspects: representational guidance approach, discussion scripting approach, and adaptive support approach~\cite{scheuer2010computer}.

The representation guidance approach encourages students to use graphical elements to represent the structure of argumentation~\cite{kirschner2012visualizing}. For example, the authors in~\cite{pinkwart2009evaluating} demonstrated that graphically representing the claim-premise structure achieved a better performance than text reading in the context of learning U.S. Supreme Court oral arguments. 
Discussion scripting approach suggests students to compare and learn from existing augmentations~\cite{fischer2013toward}. For example, show students whether their arguments are related to others' discussion and let them choose a predefined opener when drafting new arguments~\cite{huang2016group}. Adaptive support approaches~\cite{stab2014identifying, stab2017recognizing, wambsganss2019towards, scheuer2010computer, wambsganss2021arguetutor} try to provide students with adaptive recommendations, hints, and suggestions during their writing process. Adaptive support approaches are emphasized not only in the argumentation learning scope but in a broad area of reflective writing~\cite{gibson2017reflective, shum2017towards}. The feedback provided by the system could stimulate students to be self-critical, reflect on the action, and analyze their responses in writing, which helps their development of critical thinking and reasoning~\cite{schon1984reflective}. One recent work combined the representation guidance approach and adaptive support approach to guide students to learn argumentation writing~\cite{AL-CHI2020}.

However, most of them only focused on the structure of the argumentation while ignoring the fine-grained content analysis, where the persuasive strategies are reflected. In addition, few works utilize three aspects (i.e., representational guidance approach, discussion scripting approach, and adaptive support) together by providing adaptive feedback and concrete examples based on existing arguments, which is vital for argumentation learning~\cite{van2010example, schworm2007learning}.

 



\section{Formative Study}
\xingbo{Our target users are university students who constitute a considerable population in the online discussion, but most of them lack professional experience in interpersonal persuasion over the internet~\cite{lyu2020people}. In the formative study, we first conducted an online survey with 123 university students to collect their habits of using online discussion forums and requirements in improving persuasiveness in writing. For triangulation, we further conducted semi-structured interviews with five experts in debate to seek suggestions on how to address users' needs raise by our survey based on existing theories and their professional experience. Finally, we derived a set of design goals for an interactive visual system to enhance the persuasiveness of arguments in online discussions and derived a taxonomy of arguments components, based on the user survey and expert interviews.

\textbf{Participants:} We conducted an online survey with 123 university students. The participants ($age: $17$-$28$, Mean = 21.76$; 37 male, 86 female) were recruited by advertisement posts on online communities in universities. Each participant was compensated with \$5 for their time.

\renewcommand{\arraystretch}{1.5}
\begin{table}
\centering
\caption{Partial survey questions and results (Q1 - Q5). Score 1-5 are encoded using a sequential bule color scheme
\raisebox{-8pt}{\includegraphics[bb= 0 0 80 20,height=2\fontcharht\font`B]{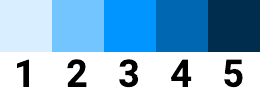}}.
The number in each chunk represents the number of participants who gave that score.}
\label{tab:survey}
\resizebox{\columnwidth}{!}{
\begin{tabular}{l|p{0.4\linewidth}|m{0.5\linewidth}}
\hline
   & Question  & Result  \\ 
\hline
Q1 & How frequently do you usually use online discussion forums, \emph{e.g.}, Reddit, Baidu forum, Quora ? (1: never, 5: very often)?                                                                                                                                                                                                                                                                 & \raisebox{-1.7\totalheight}{\includegraphics[bb= 0 0 680 10, width=0.5\textwidth]{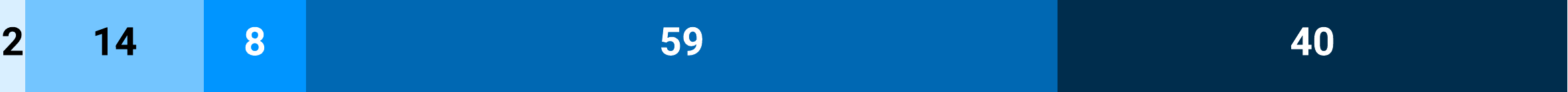}}     \\ 
\hline
Q2 & How often do you actively express opinions in the discussion on hot topics in online discussion forums? (1: never, 5: very often)                                                                                                                                                                                                & \raisebox{-1.7\totalheight}{\includegraphics[bb= 0 0 680 10,width=0.5\textwidth]{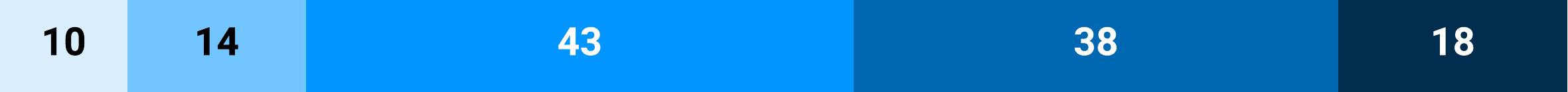}}      \\ 
\hline
Q3 & How often are you worried that the content of the published text is not convincing enough to convince others or get recognition? (1: never, 5: very often)& \raisebox{-1.7\totalheight}{\includegraphics[bb= 0 0 680 10,width=0.5\textwidth]{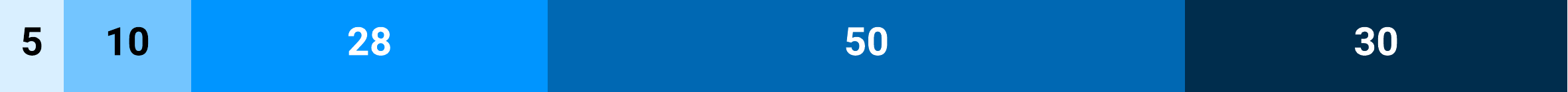}}        \\ 
\hline
Q4 & How often do you encounter situations where you need to express your views and persuade others by publishing text content? (1: never, 5: very often)                                                                                                                                                                                                      & \raisebox{-1.7\totalheight}{\includegraphics[bb= 0 0 680 10,width=0.5\textwidth]{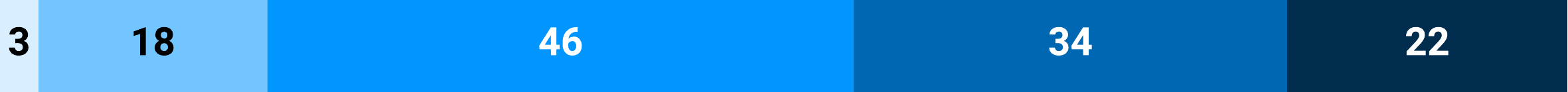}}        \\ 
\hline
Q5 & How much do you want to improve your text persuasiveness through learning so as to get more support/convince others? (1: not at all, 5: very much)                                                                                                                                                                                               & \raisebox{-1.7\totalheight}{\includegraphics[bb= 0 0 680 10,width=0.5\textwidth]{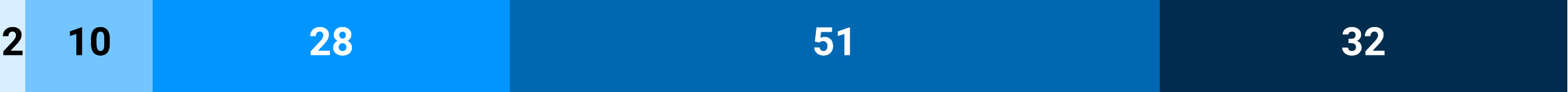}}       \\
\hline
\end{tabular}}
\end{table}

\textbf{Survey Questions:} Apart from the demographic information, the questionnaire contains (1) Likert scale questions (Q1-Q5) about participants' habits of using online discussion platforms and writing persuasive arguments, \emph{e.g.}, the frequency of their participation in the online discussion and previous experience in persuading others; (2) open-ended questions about information needed, difficulties encountered, strategies they have used, and functions envisioned for the tool when writing persuasive arguments online.

\textbf{Analysis Methods and Results:}
The questions and answers of 
participants' habits of using online discussion platforms and writing persuasive arguments) are shown in Table~\ref{tab:survey}.
For the second part, two authors analyzed the feedback from all participants using thematic analysis~\cite{braun2006using}. 
Based on participants' answers for all the survey questions, we summarize five major findings as follows. 

\textbf{F1: Participants frequently use online discussion forums and tend to express their opinions online.} It confirms that university students fit the behavioral profile of our target users. As shown in Table ~\ref{tab:survey}, nearly all participants used (121 out of 123) online discussion forums, such as Reddit\footnote{\url{https://www.reddit.com/}} and Quora \footnote{\url{https://www.quora.com/}} (Q1: $Mean=3.98, SD=1.00$) and 89\% people (110 out of 123) at least once expressed their opinions on hot topics in online discussion forums (Q2: $Mean=3.33, SD=1.11$). 

\textbf{F2: Participants want to improve the persuasiveness of their arguments online.} According to the answers from participants of Q3-Q5, we found that 
many participants (118 out of 123) were at least once worried that their arguments are not convincing enough (Q3: $Mean=3.73, SD=1.05$), and 120 out of 123 participants at least once encountered the situations to persuade others (Q4: $Mean=3.44, SD=1.03$). Most of the participants (121 out of 123) wanted to improve text persuasiveness through learning to some degree (Q5: $Mean=3.82, SD=0.97$).


\textbf{F3: Participants find it hard to come up with evidence to support their arguments.} Many participants wrote that writing or finding appropriate examples to support their opinions and make arguments persuasive is most time-consuming and challenging. P46 (F, 20) explained the difficulty of finding proper examples, \textit{``I want to find some relevant statistics to persuade others, but it is hard to find the data.''} P31 (F, 20) said \textit{``I hope to reinforce my arguments by some stories that can arouse people's emotions, but it took me very long time to make the story.''}

\textbf{F4: Participants want but consider it difficult to adapt their language style to the topic they are discussing to improve text persuasiveness.} Participants mentioned that they tried to choose a more professional language style or emotionally appealing language style to make their arguments convincing according to a specific topic. For example, P36 (F, 22) said that \textit{``I want to use some professional words and figures in my arguments to show that I have some experiences in that domain.''} P38 (F, 25) mentioned that \textit{``on some topic, people are very emotional, and it could be useful to leverage euphemistic expressions to persuade others so that they are easier to accept my opinions. However, I am a logical person and it is hard for me to write those emotional sentences.''}

\textbf{F5: People want but are not sure how to improve the logic of their arguments.} Another aspect frequently mentioned by participants is the logic of arguments, as arguments with a clear and logical structure are more persuasive~\cite{AL-CHI2020, liu2021exploring}. P15 (M, 21) gave the response, \textit{``I think it is necessary to make the whole structure of arguments clear, especially the relationship of the claim and the premise. But I am not sure whether I am doing it well.''}

In conclusion, many people frequently use online discussion forums and often encounter situations to persuade others. Therefore, they have a strong motivation to improve their arguments' persuasiveness, but their lack of expertise and experience prevents them from convincing others or even articulating their opinions online. In summary, they mainly want to improve the persuasiveness in text by learning from examples, adapting their language styles to given topics, and reflecting on the logic structure. 

\subsection{Expert Interview} \label{ssec:expert_interview}

We conducted semi-structured interviews with five experts who have rich experience in debate to seek suggestions on addressing users' needs. Debating experts can help us better understand the gaps between the current practices of the lay public and experts, as well as how to bridge the gaps for ordinary users. They also have more experience utilizing existing persuasive theories to write persuasive arguments.

\begin{figure}[htbp]
\centering
\includegraphics[width=0.6\textwidth]{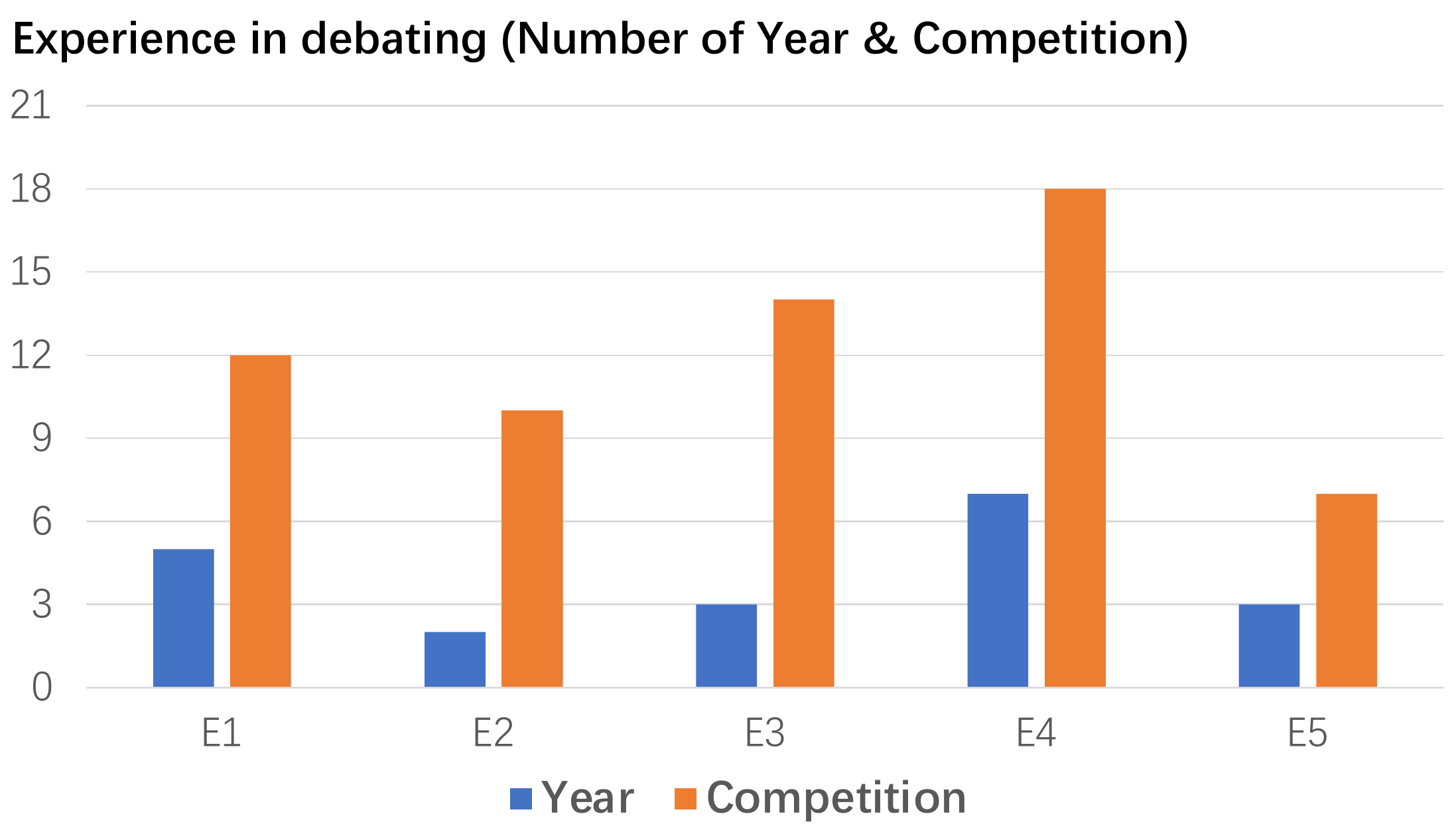}
\caption{Experience of experts in debate.}
\label{fig:expertInfo}
\end{figure}

\textbf{Participants:} Our five interviewees (three males and two females) are members of the debate teams in four different universities who have rich debate experience and are good at persuading people. All experts have theoretical knowledge and practical experience in debate. Their expertise (Figure~\ref{fig:expertInfo}) in debating ranged from 2 years to 7 years with a mean of 4 years ($SD=1.78$). All experts have participated in several debate competitions 
at the university or national level ($Mean=12.20, SD=3.01$); two of them (E1 and E2) are captains and have won the national debate competitions and used forum discussions for debate preparation.

\textbf{Interview Protocol:} Each interview lasted about an hour. We interviewed one expert at a time with open-ended questions about their experiences in persuading people, suggestions on how to address users' needs, and considerations on how to adapt existing persuasion theories \cite{tan2016winning, durmus2019modeling}. 
For example, \textit{``How do you prepare your arguments for persuading others?''}; \textit{``What strategies do you usually use to make your content more persuasive?''}; \textit{``What are your suggestions if we design a computer-aided tool for helping ordinary users to write more persuasive arguments in an online discussion?'' }; \textit{``How existing persuasive theories can be adapted to support the online discussion setting?''}, and so on. We also asked the experts follow-up questions according to their answers and specific instances of their debating competitions that they had experienced.

\textbf{Analysis Methods and Results:}
We took notes and recorded audios during each session throughout the interviews. Two authors analyzed the notes and interview transcripts and organized them based on their responses to findings (F3- F5) from the users' survey.

First, in terms of F3 (i.e., participants think it hard to find evidence to support their arguments), experts mentioned the importance of providing abundant examples in arguments. E2, E4, and E5 appreciated the strategy of using examples that fit the context and said that they usually need to learn from examples while preparing persuasive arguments, e.g., searching examples online or from books. Therefore, they thought it is essential to provide enough examples of persuasive arguments for users to learn from and make use of.
Second, they considered the different language styles in F4 (i.e., people want to adapt their language style to the topic they are discussing to improve text persuasiveness) can be mapped to different compositions of persuasive strategies (e.g., \textit{logos}, \textit{pathos}, \textit{ethos}) in persuasive theories (e.g., Aristotle's theory of persuasion). If there are more \textit{``logos''} (e.g., figures and reasoning) and \textit{``ethos''} (e.g., speakers' credibility or authorized language) in the arguments, the language style would be more professional; and if there are more \textit{``pathos''} (e.g., emotionally appealing expressions), the language style would be more evocative and affecting.
They emphasized that the composition of different persuasive strategies might vary as the topics and audience background differ. They explained that some audiences are more likely to trust objective data while others might be attracted to personal anecdotes or stories, even the same person may be persuaded using different ways under different topics. In addition, E3 and E5 stressed the importance of repeatedly improving arguments based on the reflection on one's arguments or the comparison with other arguments. For example, they usually ask their team members to comment on their arguments or discuss them by treating each other as the opposite side.
Third, they talked about the logic of (F5) persuasive arguments and emphasized the need to use logical structures to string together different examples and objective data. For instance, E5 mentioned that the claim would be less effective if there were insufficient supporting premises.

\subsection{Design Goals of the System}
\label{ssec:designgoals}
Based on the survey and interview results, we derived the following four design goals for a computer-aided system to facilitate users to write persuasive arguments in online discussions.

\textbf{DG1: Provide examples of high-quality persuasive arguments fitting to the topic and support filtering by persuasive strategies.} 
From the needs-finding survey, we found that more than half of the users referred to others' examples when writing persuasive arguments (F1). In particular, participants pointed out that they usually spent their time finding suitable examples and imitating others' writings. The expert interviews also confirm that providing abundant examples that fit the topic and demonstrated success in persuading others is beneficial for users to learn from and improve accordingly. Previous works also pointed out example-learning~\cite{schworm2007learning, van2010example, leavitt2017upvote} is an efficient way to learn argumentation skills. In addition, as mentioned in F1, some participants wanted to find statistics, and others wanted to find real stories. Therefore, the system should support filtering examples with different persuasive strategies.

\textbf{DG2: Display the composition of different persuasive strategies in the user's input and compare it with other well-accepted persuasive arguments.}
According to the survey, many participants wanted to adapt their original content to the corresponding language style of the target topic (F2). More importantly, experts mapped the language styles mentioned by users into different compositions of persuasive strategies (e.g., \textit{logos}, \textit{pathos}, \textit{ethos}) in persuasive theories (e.g., Aristotle's theory of persuasion).
In addition, experts also emphasized the importance of repeatedly improving arguments based on reflection on one's own arguments and comparing them with other persuasive arguments. Moreover, cognitive dissonance theory~\cite{festinger1962cognitive} showed that people would become uncomfortable and introspective when their existing knowledge conflicts with the information presented. Therefore, the system should display the composition of different persuasive strategies in the user's input and compare it with other persuasive arguments.

\textbf{DG3: Show the logical structure of the arguments.}
In the survey, participants mentioned the structure of the arguments might affect the persuasiveness (F3). Experts also pointed out, \textit{``Intuitive display of the logical structure is important for writing because I can have a clear grasp of what people have written.''} For example, the visualization of logical structures based on the relationship between claims and premises is helpful to check whether there are any unsupported claims~\cite{AL-CHI2020}. 

\textbf{DG4: Offer a web-based interface and visual augmented feedback for the arguments writing.}
Most participants mentioned that they hope there would be a system as intuitive as Grammarly\footnote{\url{https://app.grammarly.com/}}, a commercial website that helps grammar checking. They said that one of the key features of Grammarly is in-situ and visual augmented feedback. This requirement echos one previous research~\cite{AL-CHI2020}.}

\xingbo{\subsection{Taxonomy of Argument Components}
To address the design goals and guide the generation of feedback on persuasive strategies and logic structure, we then constructed a taxonomy of argument components, including persuasive strategies based on experts' suggestions and existing theories of persuasion. We discussed existing persuasive theories with experts, mainly including the classical persuasive strategies of Aristotle (i.e., \textit{ethos}, \textit{logos}, \textit{pathos})~\cite{rapp2011aristotle} and the extended persuasiveness measurements proposed in the context of students' essay writing (i.e., \textit{specificity}, \textit{eloquence} \textit{evidence}~\cite{carlile-etal-2018-give}. According to their experience, strategies or metrics that are impractical or semantically ambiguous are not easy to learn for lay users. We then did not adopt \textit{specificity} and \textit{eloquence} from Carlile's work~\cite{carlile-etal-2018-give}. We agreed to keep the \textit{evidence} as this is easy to understand and can satisfy users' needs for concrete examples. We also discussed the structure of the arguments with experts based on existing argumentative discourse structures in persuasive essays~\cite{Stab14-eassy}, and we decided to focus on the claim-premise relationship as in previous research~\cite{AL-CHI2020}. We present our final taxonomy of argument components as follows.

\begin{figure}[htbp]
\centering
\includegraphics[width=0.5\textwidth]{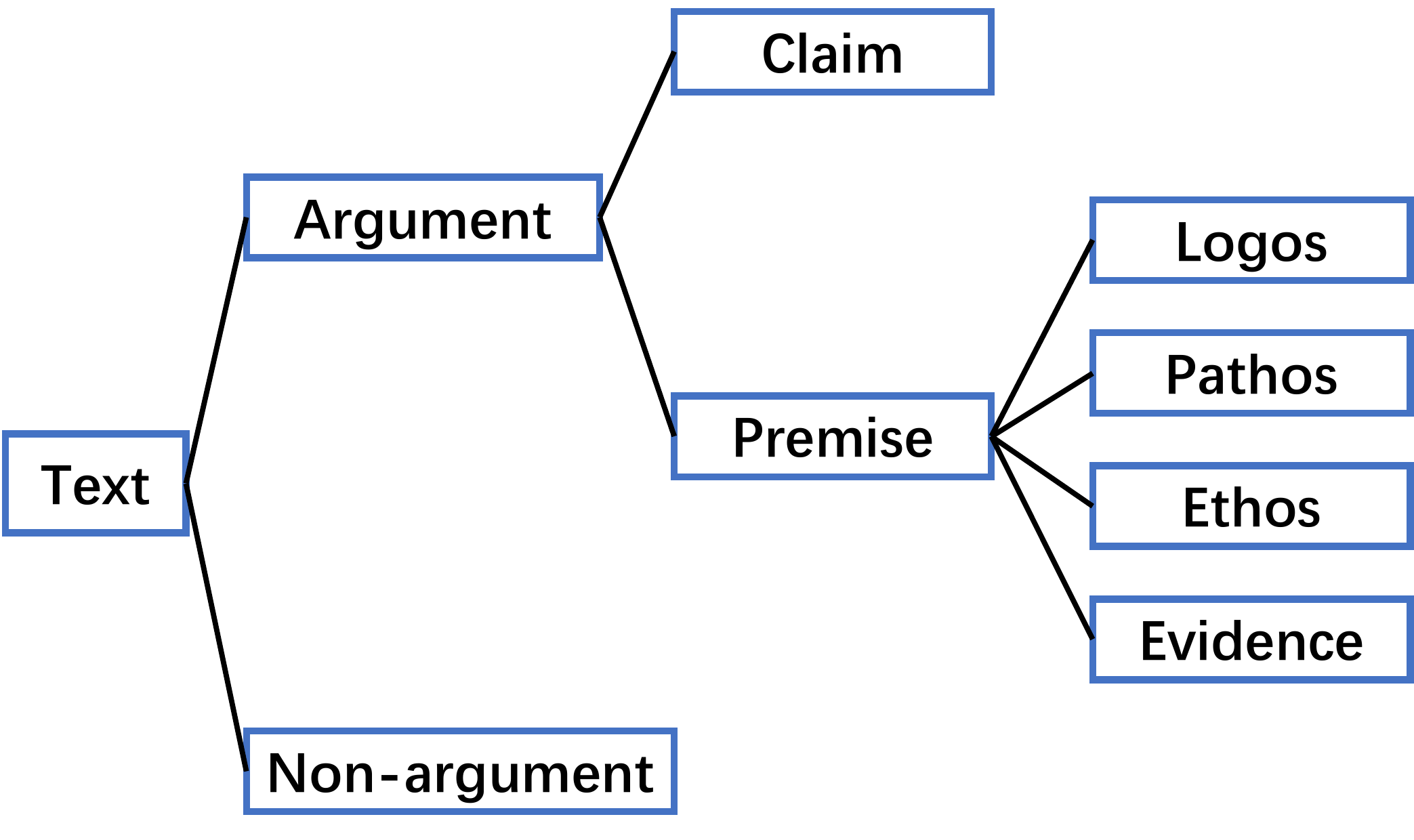}
\centering
\caption{A taxonomy with two types of argument components (claim and premise) in the argument and four persuasive strategies (i.e., \textit{logos}, \textit{pathos}, \textit{ethos}, and \textit{evidence}) in the premise.}
\label{fig:labelStructure}
\end{figure}


As shown in Figure~\ref{fig:labelStructure}, the taxonomy has two types of argument components (i.e., claim and premise) in the argument and four persuasive strategies (i.e., \textit{logos}, \textit{pathos}, \textit{ethos}, and evidence) in the premise. Each category of the label is explained below.

\begin{itemize}

\item \textbf{Argument}: The statements that include claims and premises.

\item \textbf{Non-argument}: The statements that do not have a clear position or support the overarching argument.

\item \textbf{Claim}: The statements that express the speaker's stance on something.

\item \textbf{Premise}: The statements that support a claim to persuade the audience.

\item \textbf{Logos}: A persuasive strategy that uses facts, logic, or reasoning to persuade people, such as presenting statistical results or giving logical explanations.

\item \textbf{Pathos}: A persuasive strategy that persuades people by arousing their emotions, such as telling a personal story or using emotional expressions.

\item \textbf{Ethos}: A persuasive strategy that builds the speaker's credibility by sharing some professional experiences or using credible sources for persuasion.


\item \textbf{Evidence}: A persuasive strategy that describes a concrete example. Noted that \textit{evidence} refers to the practical experience that has happened, which is not required in \textit{``logos''}, \textit{``pathos''}, and \textit{ethos}. For example, "If it is so much trouble to get dates, maintain a relationship, and not be yourself, why are you still chasing these goals" belongs to \textit{logos} but not \textit{evidence}.
\end{itemize}
}

\section{Persuasive Strategies Mining}
This section introduces how we collect corpus and label persuasive strategies and mine persuasive strategies of written texts using machine learning models based on the labeled dataset.

\subsection{Data Collection and Labeling}

\paragraph{Source Data}
We collected the source data from ChangeMyView\textsuperscript{\ref{CMV}}, which is a popular online discussion forum of Reddit\footnote{\url{https://www.reddit.com}}. 
All users can create a discussion thread under a topic by posting their claims and describing the reasons as premises. Other users attempt to change the poster's view by writing persuasive arguments.
The original poster gives $\Delta$ points to the argument that persuades him/her to be successful.
The more $\Delta$ points one reply gets, the more persuasive it is.
Thus, to build a dataset of high-quality persuasive arguments for argumentation learning, 
we filtered the replies with low $\Delta$ points under the discussion threads. \xingbo{Moreover, there are a variety of domains on CMV span economic, energy, entertainment, history, religion, politics, etc.
We chose the category ``Human Life'' for data labeling and processing since it reduces the need for professional background knowledge and contains numerous discussion posts.}
In total, we derived 164 discussion threads, covering eight topics, including abortion, dating, eugenics, immortality, marriage, parenthood, pride, and suicide.

\paragraph{Annotation of Data}
\label{ssec:label}
To provide detailed and concrete examples for persuasive strategies, we labeled the posts at the sentence level. Altogether, there are \sentence sentences. For each sentence, we first mark it as a claim or premise. If one sentence is labeled as the premise, we further label its persuasive strategy as \textit{ethos}, \textit{pathos}, \textit{logos}, or \textit{evidence}.
\xingbo{Similar to the annotation methods in the previous works~\cite{AL-CHI2020, carlile18, stab2017recognizing, liu2016effective}, three authors first labeled 10\% of the dataset and then discussed any confusing cases until a detailed coding manual was developed. For example, the annotators at the beginning were unclear on how to differentiate \textit{logos} and \textit{evidence}. After discussion, the annotators decided that \textit{evidence} refers to the practical experience that has happened, while \textit{logos} has a broader scope, including logical reasoning and what-if analysis. \emph{e.g.}, "If it is so much trouble to get dates, maintain a relationship, and not be yourself, why are you still chasing these goals" belongs to \textit{logos} but not \textit{evidence}. In addition, the annotators discussed whether showing other people's expertise can be counted as \textit{ethos}, and then decided only the expertise of the poster of the arguments can be counted as \textit{ethos}. The coding manual includes: (1) sentences are split and labeled based on the full stop while run-on sentences would be split into as multiple sentences; (2) sentences are labeled based on the context; (3) all the codes should be labeled based on the definitions introduced in Sec.~\ref{ssec:expert_interview}; (4) one sentence can only be labeled as a claim or premise; (5) the premise that have different persuasive strategies is allowed to have multiple labels; (6) \textit{evidence} refers to the practical experience that has happened in reality, while it is not required for other types of persuasive strategies (7) only the poster of the arguments can be counted as \textit{ethos}. Based on the coding manual, three annotators further labeled all the data. In our dataset, we labeled \sentence sentences, including \nonarguments non-arguments, \claim claims, \premise premises.}

\begin{figure}
\centering
  \includegraphics[width=0.6\columnwidth]{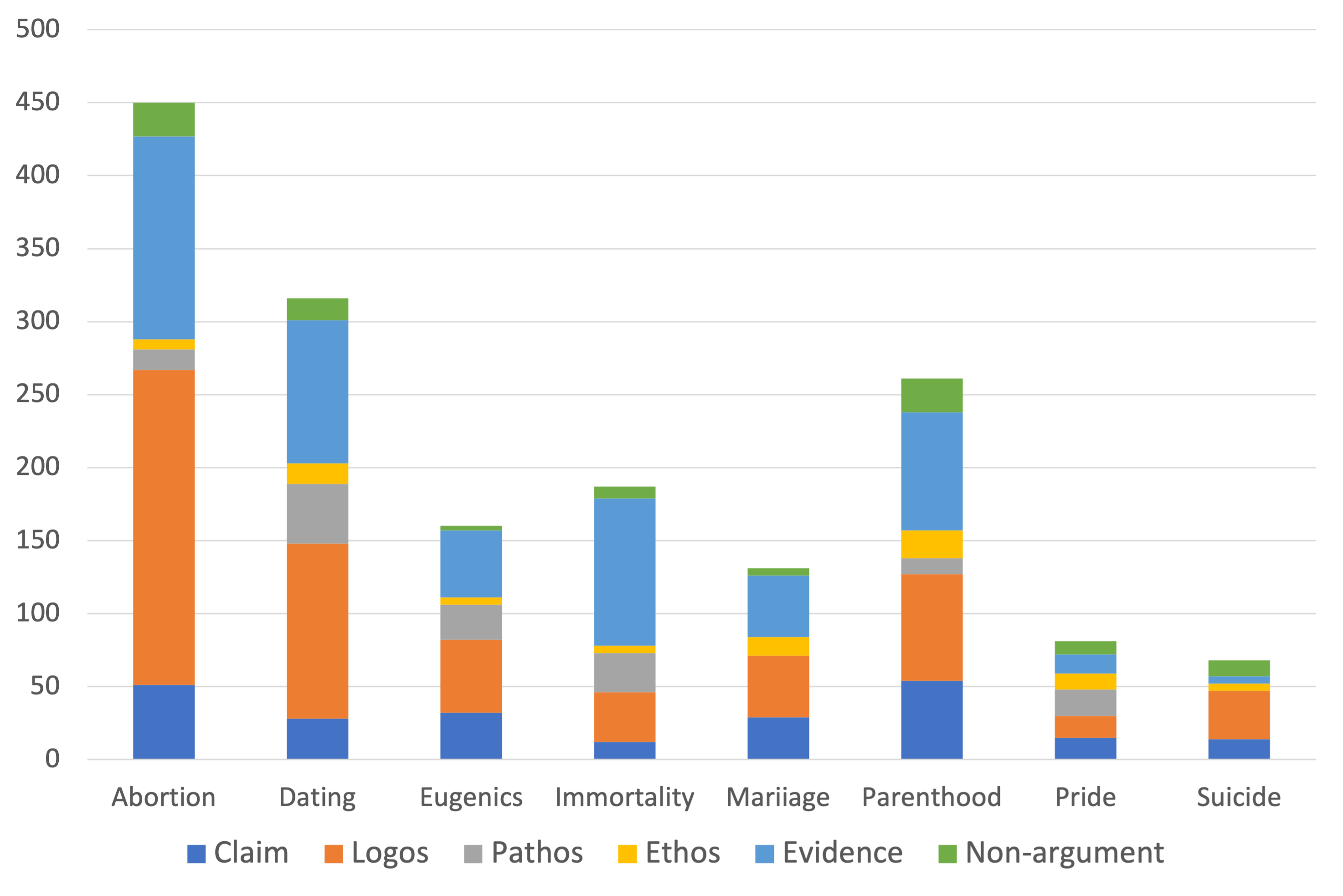}
  \caption{\xingbo{The distribution of argument components across selected discussion threads in eight different topics.}}
  \label{fig:topicComponent}
\end{figure}

\xingbo{Finally, we have labeled \logos \textit{logos}, \evidence \textit{evidence}, \pathos \textit{pathos}, and \ethos \textit{ethos}. Among 79 sentences labeled as \textit{ethos}, 39 came from the original 164 discussion threads we selected. To balance the dataset in terms of the class distribution and increase the detection accuracy of \textit{ethos}, we reviewed all the other discussion threads from the same eight topics in ``Human Life'' and collected 40 more sentences that were labeled as \textit{ethos} unanimously by the three authors. The IRR (Inter-rater Reliability) of claim, premise, \textit{logos}, \textit{pathos}, \textit{ethos}, \textit{evidence} are 0.75, 0.69, 0.74, 0.85, 0.72. We resolved conflicts through discussions among all three annotators. As for the relationship among arguments, two authors annotated the relationship between claims and premises independently. If a premise supports a claim, the corresponding claim-premise pair is labeled 1; otherwise 0. We achieved 0.78 Cohen's Kappa agreement for the relationship labeling and resolved the conflicts by discussing with a third author. The labeled result (a 0-1 array) was then used as the input of the detection model. The details of the derived dataset are shown in Figure~\ref{fig:topicComponent}.}

\subsection{Mining Method}
\begin{table}
\centering
\caption{Details about the best model's performance on the testing set of  three argumentation classification tasks:
(1). \textit{LR} for Arguments Identification;
(2). \textit{RF} for Argument Relation Detection; 
(3). \textit{Linear SVM} for Premise Classification. 
}
\label{table:algorithmT}
\begin{tabular}{llllll} 
\toprule
\textbf{Task}                                  & \textbf{Class}    & \textbf{Precision} & \textbf{Recall} & \textbf{F1} &   \\ 
\midrule
\multirow{3}{*}{Argument Component Extraction} & Claim             & 0.73               & 0.31            & 0.43        &   \\
                                               & Premise           & 0.79               & 0.97            & 0.87        &   \\
                                               & Non-argumentative & 0.91               & 0.53            & 0.67        &   \\ 
\midrule
Argument Relationship Detection                & Support           & 0.93               & 0.80            & 0.86        &   \\
                                               & Non-support       & 0.98               & 0.95            & 0.96        &   \\ 
\midrule
Premise Classification                         & Logos             & 0.79               & 0.87            & 0.83        &   \\
                                               & Pathos            & 0.83               & 0.29            & 0.43        &   \\
                                               & Evidence          & 0.75               & 0.66            & 0.70        &   \\
                                               & Ethos             & 1.00               & 0.65            & 0.79        &   \\
\bottomrule
\end{tabular}
\end{table}
\label{ssec:algorithm}
To build the system, we need to identify the argumentation tree (as shown in Figure.~\ref{fig:labelStructure}) in users' input. This process includes the detection whether a sentence is an argument or not, the detection of the relationship among claims and premises, and the detection of four sub-categories of the premises, including \textit{logos}, \textit{pathos}, \textit{ethos}, and \textit{evidence}.
With the thriving of deep learning models (\emph{e.g.}, ELMO~\cite{peters2018deep}, BERT~\cite{devlin2018bert}, and GPT-2~\cite{radford2019language}), the feature extraction process can be automated and the performance of various natural language processing (NLP) tasks (\emph{e.g.}, sentence classification) has been significantly improved. 
Inspired from a state-of-the-art model for argument classification~\cite{reimers2019classification}, 
we use BERT as the feature extractor and train a series of machine learning models on the contextual word embedding (\emph{i.e.}, features) output by BERT on the following three tasks, including argument component extraction, argument relationship detection, and premise classification.
Thereafter, we select the model with the best performance (\emph{i.e.}, highest F1 score) of each task for our application purpose (see Table~\ref{table:algorithmT}). 
We illustrate the details of our method as follows.

\paragraph{Argument Component Extraction}
We formulate the argument extraction as a multi-class classification task. Each sentence in the document corresponds to either a claim, a premise, or a non-argumentative component.
First, we split the labeled data (\sentence sentences, including \nonarguments non-argumentative, \claim claims, \premise premises) into two stratified folds: a training set (80\%) and a testing set (20\%).
\xingbo{Each set contains approximately the same ratio of target argument components.}
After converting the sentences in the dataset into high-dimensional vector representations using BERT,
we trained classical classifiers (\emph{e.g.}, Logistic Regression (LR), Random Forest (RF), Multinomial Naive Bayes (NB), Gaussian Naive Bayes (NB), Nearest Neighbor (NN), Support Vector Machine (SVM) with linear kernel and RBF kernel, and AdaBoosted Decision Tree (DT)) to derive argument components.
\xingbo{Specifically, we used 5-fold cross-validation with a stratification to measure model performance on the training data. For the detailed performance, please refer to Appendix~\ref{appendix.arg_compo}.
Then, we selected the model with the highest average cross-validation F1 score and fitted the model to the whole training data. Finally, the model was evaluated using the testing set. The performance is shown in Table~\ref{table:algorithmT}. Since the recall of claim detection was a little low, we added a rule in the user study that if no claim is detected, the first sentence is labeled as a claim by default. With this rule, the precision, recall, and F1 scores were improved to 0.79, 0.49, and 0.61, respectively.}

\paragraph{Argument Relationship Detection}
After extracting the claims and premises,
we detected the support or non-support relationship between a premise and a claim, which is a binary classification task.
Apart from the labeled support relationship of a claim-premise pair,
we generated an equal number of negative pairs for each detected claim.
In total, we derived \xingbo{250 positive pairs and 250 negative pairs.}
Similar to the previous task, we trained and selected the best classifier \xingbo{using cross validation on the training data. 
The detailed performance of each model is summarized in Appendix~\ref{appendix.arg_relation}.
Then, the best model is trained on the whole training set and evaluated on the testing set. Its performance is shown in Table~\ref{table:algorithmT}.}
Given a claim and a premise, the resulting model predicts whether the premise supports the claim or not.

\paragraph{Premise Classification}
Based on the extracted premises, we performed multi-label classification to output the types of premises. 
In our dataset, we have labeled \logos \textit{logos}, \evidence \textit{evidence}, \pathos \textit{pathos}, and \ethos \textit{ethos}. Different from the previous two tasks, each premise can have multiple labels.
To address this problem, we reduce it to multiple binary classification tasks using a simple one-vs-all~\cite{bishop2006pattern} strategy.
Specifically, we trained a binary classifier for each label, with the premises of that class being positive examples and the premises of all other labels being the negative ones.
\xingbo{Following previous procedures of model selection for argument extraction, we selected the best model for predicting the type of premise using stratified 5-fold cross validation. Please refer to Appendix~\ref{appendix.arg_premise}  for the cross validation results.}
Then, the best model is tested on the testing set (see Table~\ref{table:algorithmT}). 

\section{Interface Design}

\begin{figure}
\centering
  \includegraphics[width=\columnwidth]{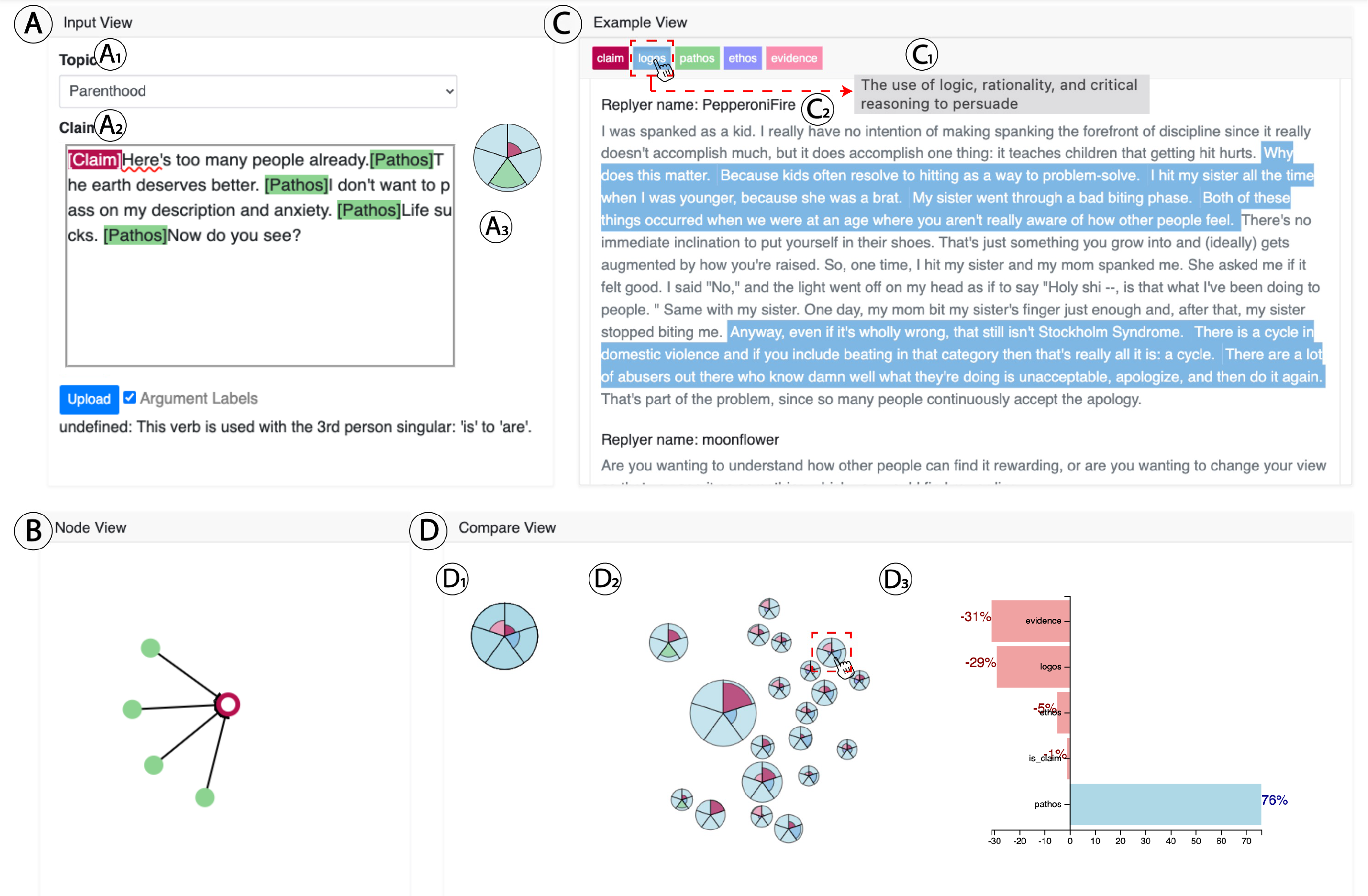}
  \caption{\xingbo{The user interface of Perusa. It contains an Input View (A) for highlighting and summarizing argument strategies applied in the user input, 
  a Node View (B) for visualizing argument structures,
  an Example View (C) for presenting a collection of persuasive argument examples, 
   and a Compare View (D) for comparing the users' argument strategies with the strategies applied by the successful arguments in the dataset.}}
  \label{fig:usage1}
\end{figure}

In this section, we describe the interface design of Persua, as well as a usage scenario, to show how our system addresses four design goals (DG1 - DG4) (Section~\ref{ssec:designgoals}).  
\subsection{User Interface}
Our system (Figure~\ref{fig:usage1}) contains four coordinated views: an \textit{Input View} (Figure~\ref{fig:usage1}\_A) that accepts a user's input and summarizes the applied persuasive strategies (DG2, DG4), a \textit{Node View} (Figure~\ref{fig:usage1}\_B) that demonstrates the structure of the user's arguments (DG3, DG4), an \textit{Example View} (Figure~\ref{fig:usage1}\_C) that provides examples of persuasive arguments fitting to the topic and supports filtering by persuasive strategies (DG1, DG4), and a \textit{Compare View} (Figure~\ref{fig:usage1}\_D) that shows the difference in arguments composition compared with other successful arguments (DG2, DG4). We use different colors to encode different argument components (at top of Figure~\ref{fig:usage1}\_C) and apply this to every view.

\paragraph*{Input View}
Input View (Figure~\ref{fig:usage1}\_A) provides a list of popular topics in ChangeMyView\textsuperscript{\ref{CMV}} for users to choose from (Figure~\ref{fig:usage1}\_A1) and accepts a user's input text (Figure~\ref{fig:usage1}\_A2). 
After the user clicks the \textit{upload} button, it marks the persuasive strategies applied in each sentence at the beginning, highlighted with the corresponding color (Figure~\ref{fig:usage1}\_A2).
To show the composition of different persuasive strategies and the claim applied in the input arguments, we use a rose chart to visualize the ratios as the user's portfolio (Figure~\ref{fig:usage1}\_A3). 
Each circular sector of the rose chart represents a persuasive strategy or a claim, as indicated by the color of the sector. The area of the sector implies the number of sentences with the specific strategy in the text, while the background sector in light-blue implies the total number of sentences. The reason for choosing this design is its intuitiveness to show proportions and easiness for comparison of portfolios, which was also used to indicate the multi-modality data (e.g., ratio of text, audio, facial) in TED videos~\cite{zeng2019emoco}. 
\xingbo{Additionally,} we also show the grammar error under the bottom of the input box.

\paragraph*{Node View}
Node View (Figure~\ref{fig:usage1}\_B) summarizes
the structure of the user's input arguments. It presents the claim-premise relationship of the input text in a node-link diagram, with each node representing a sentence. The color of a node indicates the persuasive strategies or the claim it applies. An arrow is drawn pointing a premise to the claim(s) it supports. By hovering on each node, users can see the corresponding detailed sentences.


\paragraph*{Example View}
\xingbo{Example View (Figure~\ref{fig:usage1}\_C)}
presents examples of high-quality persuasive arguments fitting to the topic and support filtering by persuasive strategies. After the user selects a topic of his/her interest, Example View presents a collection of corresponding examples (Figure~\ref{fig:usage1}\_C2) in the dataset. They are ranked by their $\Delta$ in descending order.
In addition, sentences with desired persuasive strategies can be highlighted by clicking the buttons at the top of Example View.

\paragraph*{Compare View}
Compare View (Figure~\ref{fig:usage1}\_D) helps the user compare the portfolio of the user's persuasive strategies with the portfolios of strategies applied in well-accepted persuasive arguments. First, it presents the summary of all well-accepted arguments by representing the average persuasive strategies through a rose chart (similar to that in Input View) (Figure~\ref{fig:usage1}\_D1). 
To help the user select a specific style of persuasive strategies to follow,
we project portfolios of examples (i.e., rose charts) using the multidimensional scaling (MDS) algorithm (Figure~\ref{fig:usage1}\_D2), where the position is determined by the similarity of the persuasive strategies distributions. Rose charts with similar distributions are located closer to each other. 
In addition, a bar chart (Figure~\ref{fig:usage1}\_D3) shows the discrepancy between the selected example and the user's input.
Each bar encodes a persuasive strategy and its length is proportional to the difference value by subtraction of the strategy ratio. If a premise is detected as more than one label, for example, \textit{logos} and \textit{evidence}, the system counts 0.5 for each label for comparison.
The red bars indicate the strategies that the user has less applied, with the most deficient one at the top.
The blue bars imply the strategies applied by the user exceed the selected examples.
The red color is picked for the left bars to guide users to consider more about the inadequate persuasive strategies used, which is inspired by previous research~\cite{agapie2013leading, xia2020using}.
\xingbo{Users can compare their input with an example by clicking on a rose chart. And the Example View will be scrolled to the position of the selected example.}
Meanwhile, the bar chart on the right side of Compare View will present the difference between the selected example and users' input. 

\subsection{Usage Scenario}
\begin{figure}
\centering
 \includegraphics[width=\columnwidth]{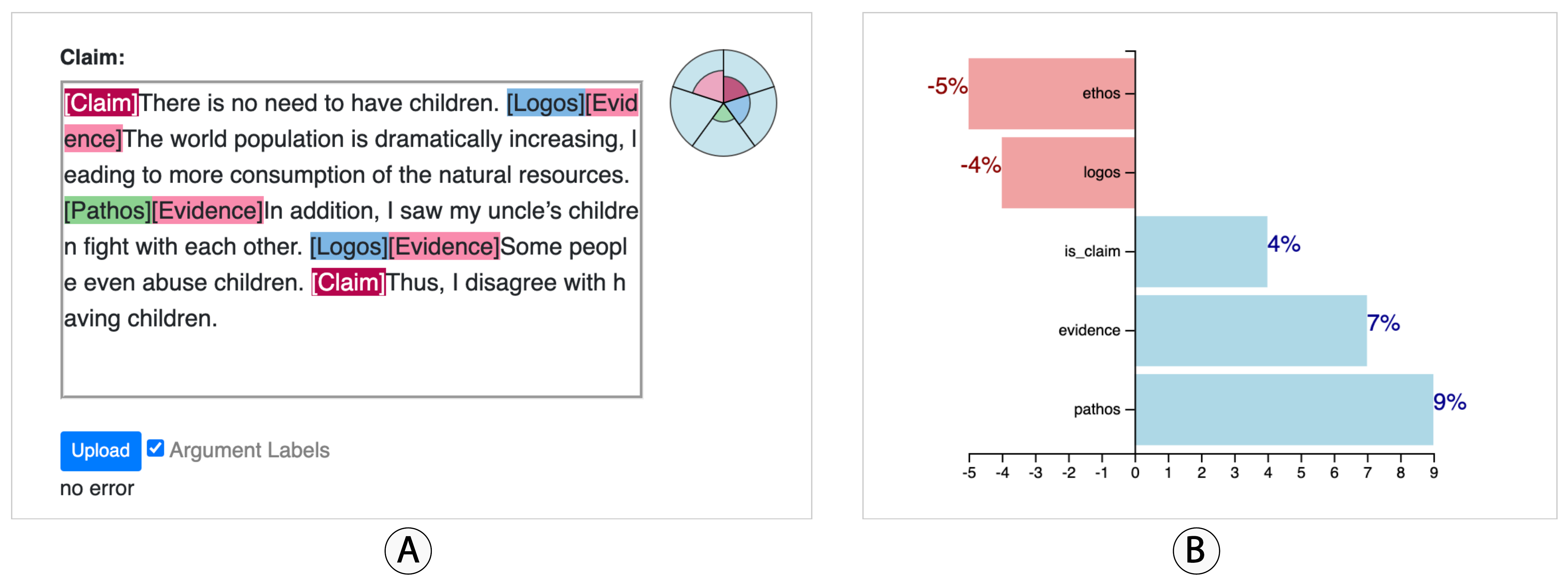}
  \caption{\xingbo{After the user refined the arguments, Persua analyzed the persuasive strategies applied by the user (A) and presented the difference between the user's arguments and those of examples from the dataset (B).}}
  \label{fig:usecase}
  \vspace{-2mm}
\end{figure}

We describe David, a user who actively participates in online discussions. Recently, he saw a post about parenthood, which encouraged people to have children. He disagreed with the post and wrote down his arguments to persuade people who posted it. However, he was not satisfied with the persuasiveness of his arguments and wanted to improve it. Then, he referred to the Persua.
He first selected the topic ``parenthood'' in Input View (Figure~\ref{fig:usage1}\_A1) and uploaded the paragraph he has written. Then, he examined the feedback in Input View (Figure~\ref{fig:usage1}\_A). 
As shown in Figure~\ref{fig:usage1}\_A, he noticed that each sentence has been labeled with one or more persuasive strategies or ``claim'' in Input View, such as \textit{claim} in front of the first sentence, \textit{pathos} before the other sentences. Then he found the logical structure of his text was shown in Node View (Figure~\ref{fig:usage1}\_B) and confirmed that no dangling claim without support.
To improve the persuasiveness of his arguments, he wanted to learn from well-accepted examples and checked the 
cluster of glyphs in Compare View (Figure~\ref{fig:usage1}\_D). He found that some glyphs on the left were clearly red and glyphs on the right were more blue. This indicates good examples have different persuasive strategies, \emph{i.e.}, arguments on the left have more \textit{claims} and \textit{evidence}, while arguments on the right have more \textit{logos}. By checking the bar chart (Figure~\ref{fig:usage1}\_D3), which summarized the difference between David's persuasive strategies (Figure~\ref{fig:usage1}\_A3) with the average persuasive strategies of all the examples (Figure~\ref{fig:usage1}\_D1), he noticed two long red bars on the left, labeled with \textit{evidence} and \textit{logos}, which meant that comparing with the average persuasive strategies components, David's arguments lacked \textit{evidence} and \textit{logos}. Then he clicked one glyph as highlighted in Figure~\ref{fig:usage1}\_D2, which had a large portion of blue (\emph{i.e.}, \textit{logos}). Then the corresponding arguments of this glyph were shown in Example View with \textit{logos} highlighted (Figure~\ref{fig:usage1}\_C2). He read the two highlighted episodes to get inspiration for adding \textit{logos} in the arguments. The sentence~\textit{``My sister went through a bad biting phrase''} reminded him of an experience that he saw naughty kids always fighting with each other. In addition, the second highlighted example made him imagine there were some people who abused children. Then he rephrased these sentences and added to his arguments. He uploaded and the system feedback was shown in Figure~\ref{fig:usecase}. He found that \textit{logos} ratio increased from $-31$ to $-4$ and \textit{evidence} become positive, and finally got satisfied with his writing.
 By adding some other types of sentences, David enriched the overall content based on exploring and learning in our system. 

\section{Evaluation}
We conducted a between-subject user study to evaluate the usefulness and usability of Persua in helping online forum users improve the persuasiveness of written texts.

\begin{figure}[h]
  \centering
  \includegraphics[width=0.75\linewidth]{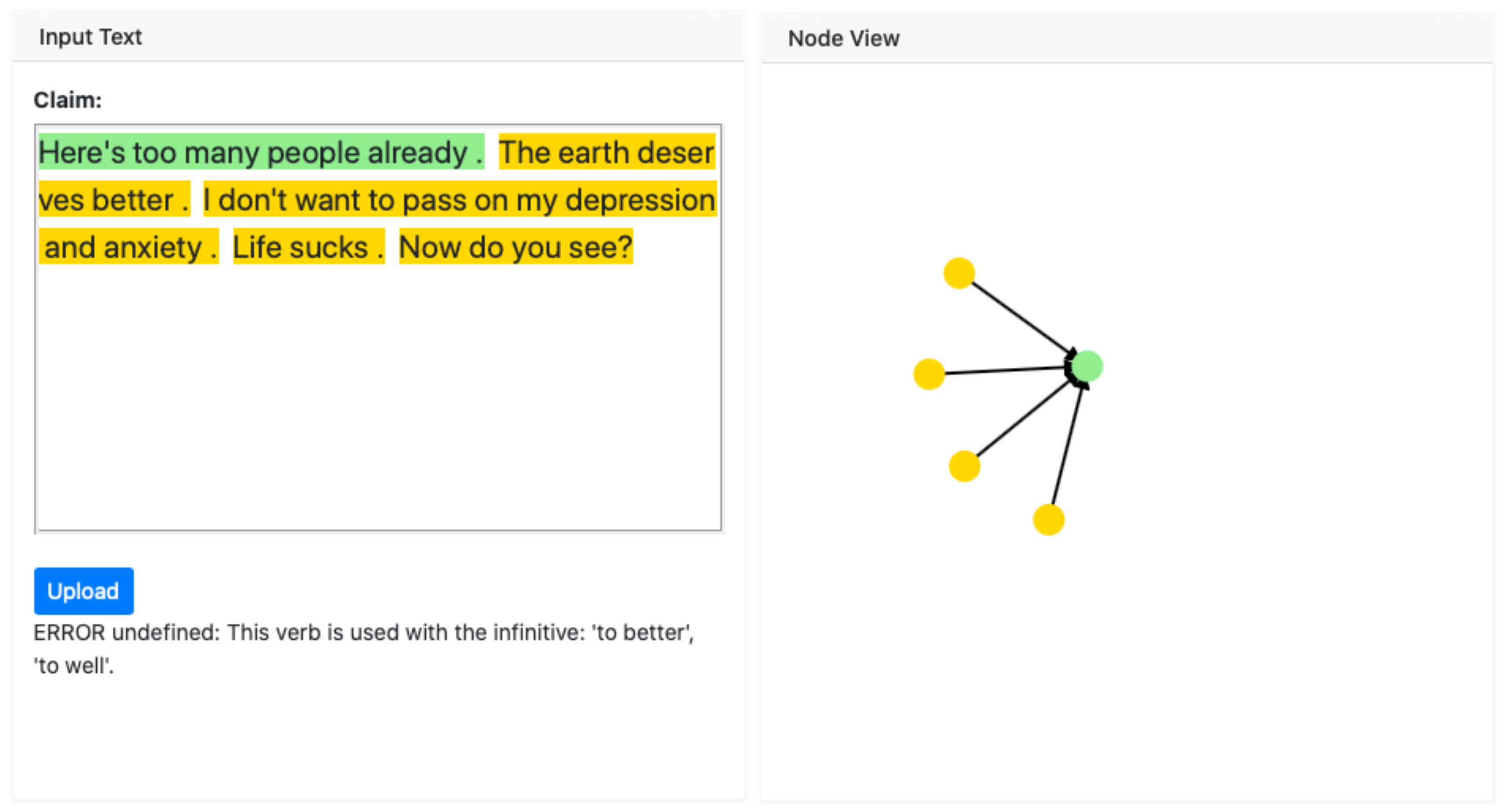}
  \caption{The interface of the Baseline system.}
  \Description{}
  \label{fig:Baseline}
\end{figure}
\subsection{Experiment Design}
\textbf{Baseline System:}
\xingbo{We built the Baseline system referring to a recent work ~\cite{AL-CHI2020}, which has a similar goal---improving students' argumentation skills. As shown in Figure~\ref{fig:Baseline},
we implemented the main feature of AL, demonstrating the claim-premise relationship using a node-link diagram, to construct the baseline system for testing. For some other features in the previous work, \emph{e.g.}, the readability score, we replaced them using similar features in Persua, \emph{e.g.}, showing the grammar errors below the input text. We did not reproduce all features since we focus more on comparing the helpfulness with or without showing fine-grained persuasive strategies, rather than comparing two systems.}


\xingbo{\textbf{Participants:}
We recruited 36 participants ($male:20, female:16, age: Mean = 24.08, SD = 3.44$) from universities through mailing lists and social networks. Participants are from nine couturiers and districts (Mainland China: 23, South Korea: 4, India: 2, Kazakhstan: 2, Hong Kong: 1, Indonesia: 1, Russia: 1, Syria: 1, Turkey: 1). Their average experience on online discussion forums is $4.28$ years (SD = 2.47). Participants were divided into two groups: 18 people in the Baseline group (A1-A18, $male:11, female:7$ age: $Mean = 23.89, SD = 3.46$, years using online discussion forums: $Mean = 4.06, SD = 2.49$, Mainland China: 12, South Korea: 2, India: 1, Kazakhstan:1, Russia: 1, Syria: 1) and 18 people in the Persua group (B1-B18, $9$ male, $9$ female, age: $Mean =24.28, SD = 3.51$, years using online discussion forums: $Mean = 4.50, SD = 2.50$, Mainland China: 11, South Korea: 2, India: 1, Kazakhstan: 1, Hong Kong: 1, Indonesia: 1, Turkey: 1). Participants signed the consent form before the user study and agreed to discuss the topics about abortion and parenthood, which are used in the evaluation tasks. Each participant received $\$8$ as compensation for participating in a 90-minute study session.}

\textbf{Experiment Procedures:}
\begin{table}[!b]
\caption{Tasks description.}
\label{tab:tasks}
\begin{tabular}{@{}ll@{}}
\toprule
Task 1 & \begin{tabular}[c]{@{}l@{}}Write a paragraph about the topic ``Abortion''. \\ Try to persuade the person who published the following claim by taking the opposite side.\\ {[}Claim: I'm pro-life, and I believe that abortion is essentially murder.{]}\end{tabular} \\ \midrule
Task 2 & \begin{tabular}[c]{@{}l@{}}Refine a paragraph about the topic ``Parenthood'':\\ ``\textit{Here's too many people already. The earth deserves better.}\\ \textit{I don't want to pass on my depression and anxiety. Life sucks. Now do you see?}''\\ Try to make it more convincing to persuade people who published the following claim.\\ {[}Claim: I don't understand why people don't want to have children.{]}\end{tabular} \\ \bottomrule
\end{tabular}
\end{table}
The procedure of this study contains three sessions. First, participants were briefly introduced to the background of the project and the system interface for about five to ten minutes. \xingbo{We introduced the definitions of different labels using the tooltips in the system as shown in Figure~\ref{fig:usage1}\_C1.} Second, participants were asked to complete two tasks, as shown in Table~\ref{tab:tasks} in a think-aloud way.
These two tasks tried to test system usefulness on two common scenarios.
Task 1 asked participants to draft a paragraph of arguments from scratch to persuade a person who holds an opposite opinion under a given topic;
Task 2 required participants to refine a paragraph under a given topic.
Third, participants were asked to fill a post-study questionnaire with 7-point Likert scale questions in Table~\ref{tab:questions} and two open-ended questions: (1) which view is the most useful? Which view is the least useful? Why? (2) Any suggestions on improving the system?.

\textbf{Measures}
\xingbo{We followed the evaluation pipeline of interactive systems in Weibelzahl's work~\cite{weibelzahl2001evaluation} to assess our system from the following three aspects:}
system usefulness, visual design \& interactions, and system usability. The questionnaire (Table~\ref{tab:questions}) was derived based on existing research~\cite{xia2019peerlens}, to test the usefulness, visual design and interactions, and usability. We ran Mann-Whitney U test on each rating since it is a non-parametric test ($n_1=n_2=18$) on unpaired ordinal data~\cite{mcknight2010mann}. \xingbo{Since people's trust is an important metric in an AI-infused system~\cite{yang2020re}, we also evaluated learners' perceived accuracy of the information in Persua. This is based on their verbal feedback towards the algorithm's performance in the user study and their suggestions for the system in the questionnaire.}

The two systems logged all submissions of arguments. Apart from users' self-ratings toward the system and their qualitative feedback during the user study, we also invited two experts (E1, E2) to rate the quality of the written texts. The scores were given on a Likert scale from 1 to 5 points (1: not persuasive, 5: very persuasive). As participants may revise the written texts more than one time for each task, we used the first-time and last-time submissions to represent the persuasiveness before and after using the systems. We then took the average score of the two experts for each submission. To eliminate the influence of users' inherent argument skills, we only focused on the change of the experts' ratings on arguments before using the system and after using the system. 
\xingbo{In particular, the criteria are
(1) overall logic: arguments that have a direct relationship between the evidence and the conclusion drawn receive a higher score, vice versa;
(2) connection: arguments that have apparent conjunctions or connections between arguments receive a higher score, while jumpy thinking or a list of points that are not closely aligned receive a lower mark;
(3) support: the arguments that are factually true or supported by evidence receive a higher score, vice versa. 
(4) comprehensiveness: answers that take more perspectives of a problem into account
receive a higher score than those that consider about only one side;
(5) qualification: if there are qualified words such as ``in my opinion'', it will score more than absolute terms; similarly, answers that generalize their ideas to the whole of society or others receive a lower mark.}	

\begin{table}[]
\caption{ Our questionnaire focuses on 3 aspects: informativeness (Q1$-$Q4), visual design \& interactions (Q5$-$Q6), system usability (Q7$-$Q9).}
\label{tab:questions}
\begin{tabular}{c|l}
\hline
Q1  & \begin{tabular}[c]{@{}l@{}}The system provides enough information to improve the persuasiveness of arguments.\end{tabular}                             \\ \cline{2-2} 
Q2  & \begin{tabular}[c]{@{}l@{}}Node View helps me improve the persuasiveness of my writing.\end{tabular}                             \\ \cline{2-2} 
Q3  & \begin{tabular}[c]{@{}l@{}}Example View helps me improve the persuasiveness of my writing.\end{tabular}                             \\ \cline{2-2} 
Q4  & \begin{tabular}[c]{@{}l@{}}Compare View helps me improve the persuasiveness of my writing.\end{tabular}                             \\ \cline{2-2}                          
\hline
Q5  & The visual designs in the system are intuitive.                                                                                                   \\ \cline{2-2} 
Q6  & The interactions in the system are intuitive.                                                                                                   \\ \cline{2-2}
\hline
Q7  & It was easy to learn the system.                                                                                                         \\ \cline{2-2} 
Q8 & It was easy to use the system.                                                                                                           \\ \cline{2-2} 
Q9 & I will use this system again.                                                                                         \\ \hline
\end{tabular}
\end{table}

\textbf{Hypothesis:} 
Example-based learning can facilitate students' learning in ill-structured domains such as argumentation~\cite{van2010example, schworm2007learning}. And showing examples of persuasive strategies were also demanded by 90\% of respondents in our need-finding survey. We thus proposed the hypotheses as follows. 

\begin{itemize}
    \item \textit{\textbf{H1:}}
     Persua is more useful than the Baseline on improving the persuasiveness of users' writing. Specifically, the improvement of the text persuasiveness between the first submission and the last submission in Persua was higher than Baseline (H1a). Moreover, participants thought that Persua provided more information to help them improve persuasiveness (H1b) and \xingbo{they submitted their arguments more times for feedback when using Persua than Baseline (H1c).}
     \item \textit{\textbf{H2:}}
     Persua is more intuitive than the Baseline. Specifically, participants had higher ratings on the visual design (H2a) and the interaction than Baseline (H2b).
     \item \textit{\textbf{H3:}}
     Participants preferred using Persua than the Baseline. Specifically,
     \xingbo{Persua was perceived better in terms of}
     ``easy to use'' (H3a), ``easy to learn'' (H3b), and ``willing to use again'' (H3c).
\end{itemize}





\subsection{Results and Analysis}
\xingbo{
Overall, task performance demonstrates that the improvement of persuasiveness between the first submission and the last submission in Persua is higher than that in the Baseline. In addition, there is no significant difference in the ratings between Persua and Baseline on the visual design intuitiveness and interaction intuitiveness. Although Baseline is reported as easier to learn and use due to its simple functions and less informative views compared to Persua, participants were more willing to use Persua in the future when they write arguments online.


\begin{figure}[h]
  \centering
  \includegraphics[width=0.55\linewidth]{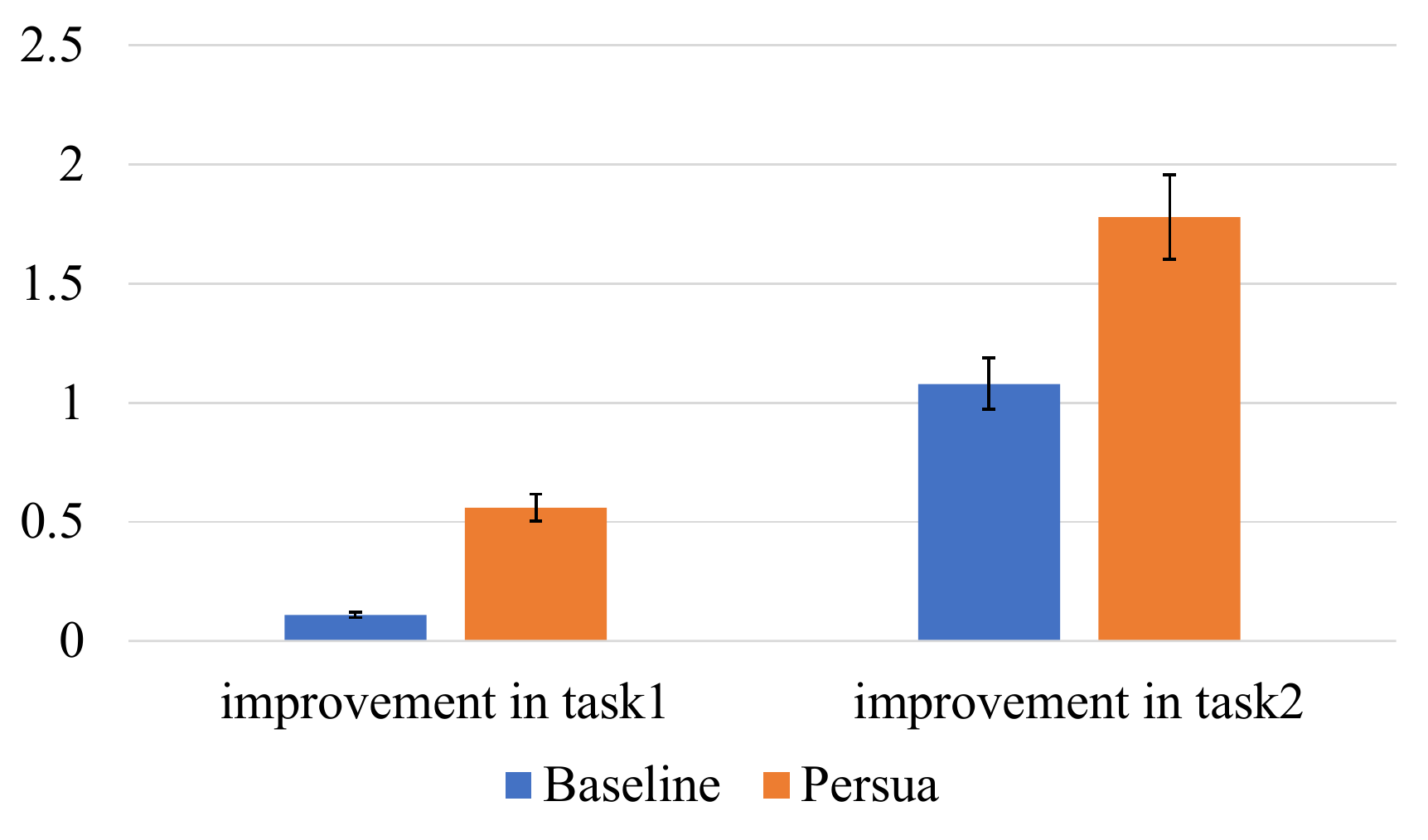}
  \caption{\xingbo{Improvements between the first submission and the last submission on the persuasive scores for both tasks and systems using a 5-point Likert scale.}}
  \Description{}
  \label{fig:tasks}
\end{figure}

\begin{figure}[h]
  \centering
  \includegraphics[width=\linewidth]{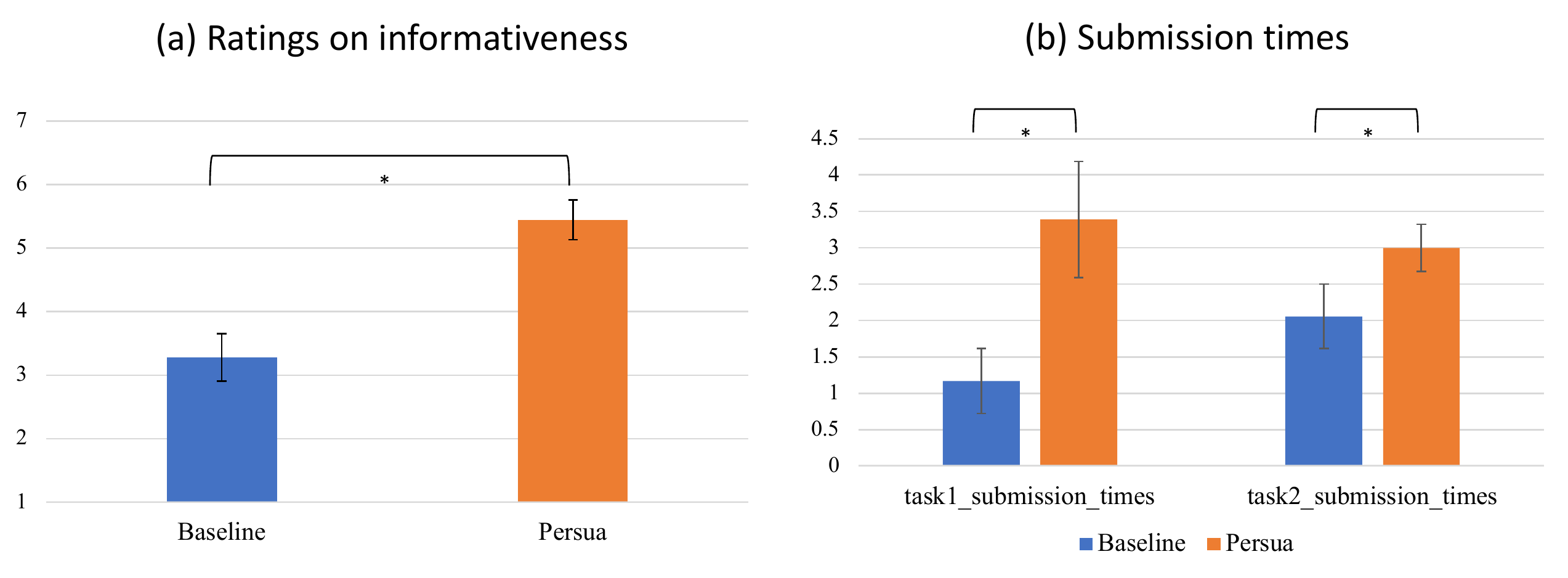}
  \caption{(\xingbo{a) Participants' ratings on informativeness of Baseline and Persua on a 7-point Likert scale ($*:p<.05$). (b) Submission times in both tasks and systems.}}
  \Description{}
  \label{fig:informativeness}
\end{figure}

\begin{figure}[h]
  \centering
  \includegraphics[width=0.90\linewidth]{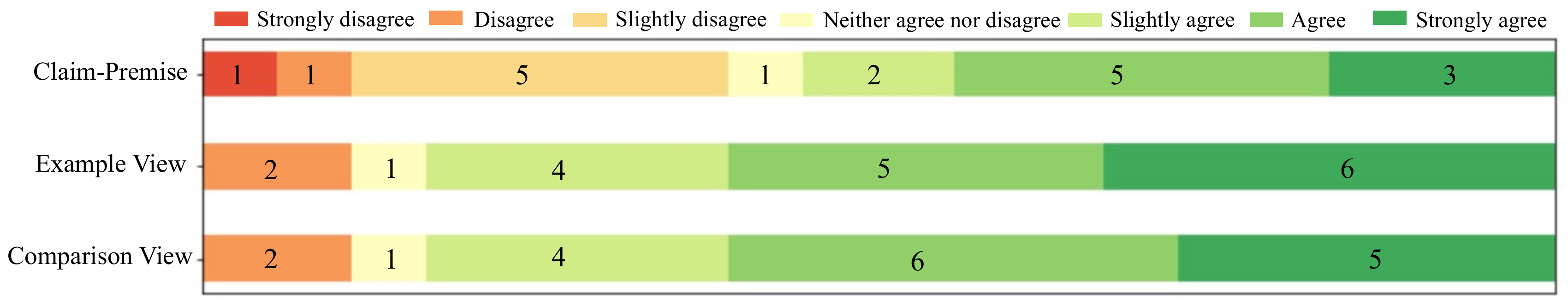}
  \caption{\xingbo{Participants' ratings on the usefulness of Node View, Example View, and Compare View on a 7-point Likert scale ($*:p<.05$).}}
  \Description{}
  \label{fig:view_ratings}
\end{figure}

\textbf{System Usefulness}
The task performance shows that Persua is more helpful for users to improve the persuasiveness in their arguments more than Baseline.
First, all improvements in persuasive scores between the first submission and the last submission are higher in Persua than Baseline in Task 1 and Task 2, and there is a marginally significant difference in Task 2. As shown in Figure~\ref{fig:tasks} a, on Task 1 which is writing a paragraph from scratch, the score change in Persua ($Mean = 0.56, SE = 0.20$) is higher than the score change in Baseline ($Mean = 0.11, SE = 0.09$). The Mann-Whitney U test shows that $p = 0.134 > 0.05$. On Task 2, refining a paragraph, the score change of Persua ($Mean = 1.78, SE =0.24$) is higher than that of the Baseline ($Mean = 1.08 , SE = 0.25$). The Mann-Whitney U test shows that there is a marginal significance with $p = 0.06 > 0.05$, though H1a rejected.
In addition, the results showed that Persua provided more information on how to improve the argument's persuasiveness and encouraged participants to edit and submit more times for feedback. According to the participants' ratings on the system informativeness, Persua ($Mean = 5.33, SE = 0.47$) scores significantly higher than Baseline ($Mean = 3.33, SE = 0.50$). The result of Mann-Whitney U test showed that $p =  < 0.05$, H1b supported.
We also noticed that participants in Persua edited more times and submitted more times than those who used the Baseline. The submission times of Baseline is $Mean = 1.17 , SE = 0.38$ in Task 1 and $Mean = 2.06 , SE = 0.54$ in Task 2, while the submission times of Persua is $Mean = 3.39 , SE = 3.38$ in Task 1 and $Mean = 3.00, SE = 1.37$ in Task 2. The followed Mann-Whitney U test showed that $p = 0.008 < 0.05$ in Task 1 and $p = 0.047 < 0.05$ in Task 2, H1c supported.

In particular, participants utilized multiple views in Persua in different ways to improve their persuasiveness. To understand which view contributed more to the current result, we checked participants' ratings on the usefulness of Node View, Compare View, and Example View, which is shown in Figure~\ref{fig:view_ratings}. The result indicated that participants thought Compare View and Example View are more valuable than Node View (i.e., claim-premise structure). Furthermore, we observed that 13 out of 18 participants in the Persua group used Compare View to compare their strategy composition with the average persuasive strategy composition and added more particular persuasive strategies they lacked by referring to Example View. For example, B16 noticed from Compare View that the arguments she wrote had less \textit{logos} than the average strategy composition, and then she learned more \textit{logos} from Example View and further enriched her arguments by adding more \textit{logos}. In addition, 10 out of 18 participants used Example View in Persua to inspire them before writing arguments. For example, B13 checked all the \textit{claims} in Example View to quickly grasp the opinions of all persuasive arguments; B14 said he had less knowledge about ``Abortion'' and did not know how to start at the beginning. After going through arguments in Example View, he was inspired by some examples relevant to ``animal'' and thought babies are like animals, and if the mother cannot provide a good life, the baby would be miserable like some animals. A few participants only used the portfolio of persuasive strategies of their own input to reflect on how to improve instead of comparing themselves to others. 
    
The verbal feedback from participants in both Baseline group and Persua group showed us why Node View was not that helpful. In the online discussion setting, the arguments are usually short, with one claim and several premises. The claim-premise relationship is relatively straightforward in this case, and thus the feedback provides limited information for users to reflect on and revise their arguments. For example, B1 said that ``\textit{There are only a few sentences, and I know their relationship.}'' B14 commented that, ``\textit{I have no problem in making the arguments a well-structured paragraph. However, I want more feedback and suggestions on the content and tone of the arguments.}'' 
Participants also reported issues that Persua can improve. For example, some participants mentioned that the summary portfolio and the bar chart are more helpful than the clustering graph in Compare View. The reason is that they could not easily find common patterns of successful arguments in the clustering graph, and they suggested adding more unpersuasive examples of arguments, which may help them differentiate successful and unsuccessful arguments from the clustering graph.} 

In conclusion, H1 is partially supported, Persua can help users improve the persuasiveness in their writing more than Baseline. Example View and Compare View in Persua contribute more to the system's usefulness.

    




\begin{figure}[h]
  \centering
  \includegraphics[width=0.55\linewidth]{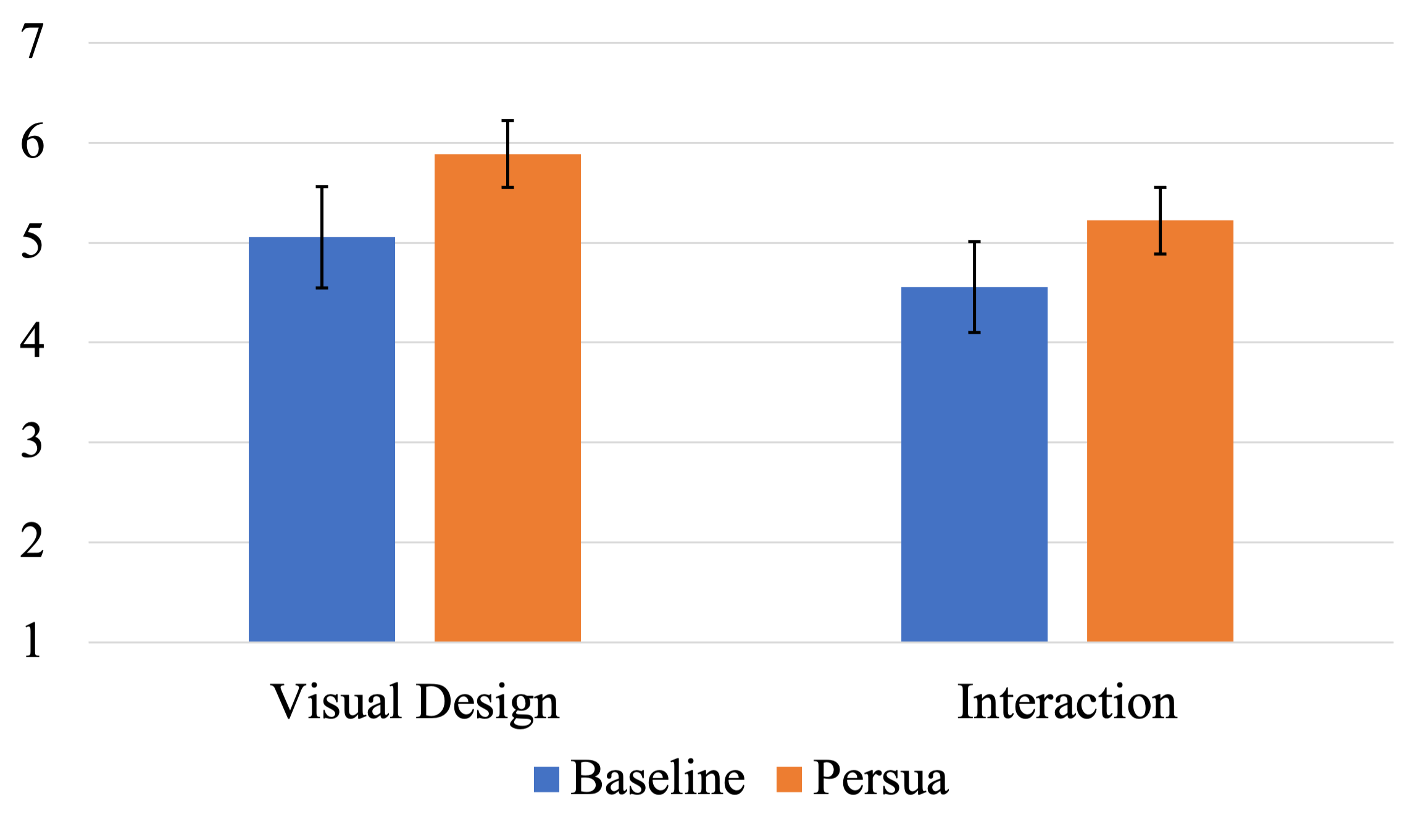}
  \caption{\xingbo{Participants' ratings on visual design \& interactions of Baseline and Persua on a 7-point Likert scale ($*:p<.05$).}}
  \Description{}
  \label{fig:visual_design}
\end{figure}

\textbf{Visual Designs \& Interactions}
Overall, participants thought visual designs and interactions in both Persua and the Baseline were intuitive, and there was no significant difference (H2 rejected). Participants' ratings on the intuitiveness of the visual designs are shown in Figure~\ref{fig:visual_design}. Persua ($Mean = 5.06, SE = 0.38$) had a similar distribution with the Baseline ($Mean = 5.89, SE = 0.27$). The result of Mann-Whitney U test shows that $p = 0.161 > 0.05$, H2a rejected. Participants' ratings on the intuitiveness of the interactions are shown in Figure~\ref{fig:visual_design}. Persua  ($Mean = 4.56, SE = 0.41$) had a similar distribution with the Baseline ($Mean = 5.22, SE = 0.33$). The result of Mann-Whitney U test shows that $p = 0.293 > 0.05$, H2b rejected. In the user study, we introduced each concept by referring its definition. Though the scores of intuitiveness in visualization and interaction of Persua are a bit lower than Baseline, it is reasonable since Persua has much more visual elements and views than Baseline. Participants also pointed out issues in visual designs that could be improved. For example, the size of the circle in Compare View was misunderstood by some participants as the power of persuasiveness. 

\begin{figure}[h]
  \centering
  \includegraphics[width=0.55\linewidth]{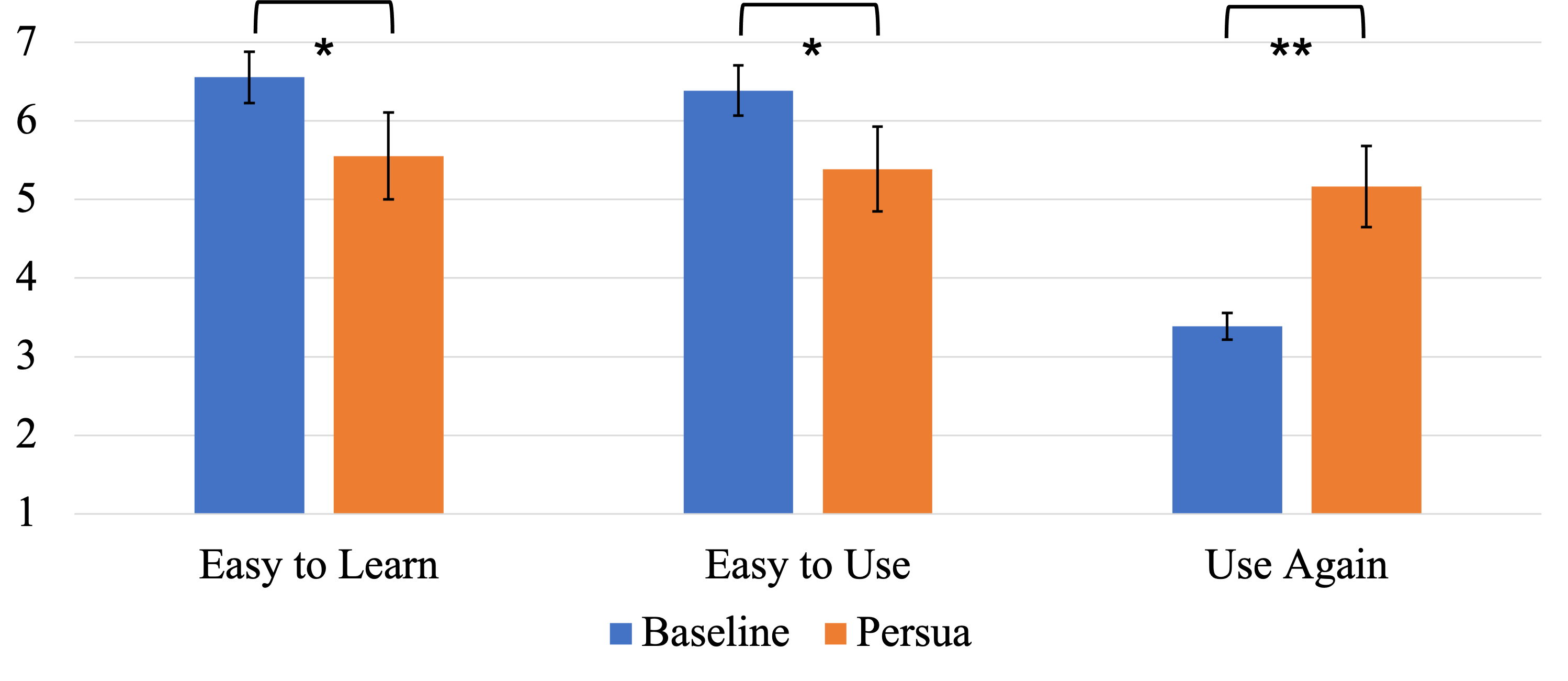}
  \caption{\xingbo{Participants' ratings on system usability (easy to learn, easy to use, willing to use again) of Baseline and Persua on a 7-point Likert scale ($*:p<.05$, $**:p<.01$)).}}
  \Description{}
  \label{fig:system_usability}
\end{figure}

\textbf{System Usability} In general, participants were significantly more willing to use Persua again than Baseline, though they rated Baseline significantly more easy to learn and use, as shown in Figure~\ref{fig:system_usability} (H3 partially supported). In both questions of easy to learn and easy to use, Persua ($Mean = (5.56, 5.39), SE = (0.35, 0.35)$) received average scores more than 5, while lower than the Baseline system ($Mean = (6.56, 6.39), SE = (0.18, 0.20)$). Mann-Whitney U test results show that $p = 0.029, .027 < 0.05$ for both former and latter cases, so H3a and H3b are rejected. Participants' ratings on willingness to use Persua again are shown in Figure~\ref{fig:system_usability}, here Persua ($Mean = 5.17, SE = 0.35$) had a higher score than the Baseline ($Mean = 3.39, SE = 0.46$). The result of the Mann-Whitney U test shows $p < 0.01$, H3c accepted. \xingbo{Most participants could grasp the meaning of \textit{logos} immediately after introduction, and around two participants in each group referred to the definition in the tooltip, and there seemed no particular difficulty for them to understand these labels. However, some participants suggested that if example arguments in Example View could be ranked based on their similarity with the given arguments, it would be easier to look for suitable materials. In addition, a few participants (B5, B8) mentioned that when they selected one persuasive strategy, e.g., \textit{logos}, in Example View, they found it tedious to scroll a lot to check the highlighted \textit{logos} in all the examples. Therefore, we need to consider how to keep the readability without losing the context of a paragraph by hiding all the other sentences.}

\xingbo{\textbf{Perceived Algorithm Accuracy}
Overall, the algorithm accuracy was acceptable by most of the participants in the user study, while it needs further improvement. During the user study, we asked for participants' feedback on the algorithm performance when they conducted the tasks. Most of the participants thought that the results seemed reasonable with their inputs. This can also be reflected by their positive scores on ``willing to use the system again'' ($Mean = (5.56, 5.39), SE = (0.35, 0.35)$). However, two of the 18 participants in Baseline (A10, A15) and two of the 18 participants (B7, B11) in Persua noticed that some of their claims were not detected. Most of the cases were that they had sub-claims which were not the first sentence. For example, A15 mentioned  ``\textit{I can have a claim for the overall paragraph, and have smaller claims to support the overarching claim with the evidences. But the system only presents the first sentence as the claim and others are classified as premises. So, it was hard for me to catch which sub-claim was well-supported and which one would need more reinforcement. If such a more fine-grained visualization was available, I think the system would be more useful.}'' These four participants replied in the questionnaire that the claim-premise detection algorithm could be improved to detect more claims and support more complex claim-premise relationship. In addition, one participant (B13) reported that the persuasive strategy detection results were different from his initial thinking. However, he tried to interpret the algorithm result by saying  ``\textit{maybe the algorithm thought that I used an emotional word, thus it detected it as \textit{pathos} instead of \textit{logos}}''. He further suggested that the algorithm should take the context into consideration. In conclusion, the current algorithm can generally support the user study, the accuracy should be further improved, and the visualization could be refined to convey the uncertainty of AI algorithms.}
\section{Discussion}
This section discusses the design considerations we learned from the study and the limitations that we can address in future work. These findings could help design an interactive system based on collaboratively generated data to write persuasive arguments and contribute to a more effective and civil online discussion environment.
\subsection{Design considerations}
We observed several interesting usage patterns of Persua during the experiments, from which we derived a set of design considerations for the future development of intelligent systems to improve the persuasiveness of writing in online discussion. 

\textbf{Encourage creative thinking by showing detailed premise types.}
The different premise types and persuasive strategies (\emph{i.e.}, \textit{logos}, \textit{pathos}, \textit{ethos}, and \textit{evidence}) work as a stimulus to encourage people to consider different methods to persuade other people.
During the experiment, participants commented that the definitions of the different premise types given in Example View are very beneficial. For example, B12 said ``\textit{These labels listed in Example View works like a checklist, which reminds me to think more.}. \textit{I am not good at using emotional expressions, and the \textit{pathos} strategy makes me think of other possibilities to persuade people.}''  Like the six-hats thinking method~\cite{de2017six}, which encourages people to have both critical thinking and creative thinking, the various types of premises guide people to think creatively and produce more supportive sentences for their claims. 

\textbf{Enable users' autonomy on how to use examples at different stages of arguments writing.}
The initial intent for us to design Example View is to provide users with a handy resource from which they could search for application instances of a specific type of premise, \emph{e.g.}, \textit{logos}. Indeed, participants have various ways to utilize the examples provided for argument writing. Some users try to use examples as the starting point to get inspiration for the writing. Other participants in the Persua group proceeded to Example View after checking Compare View to find materials that they could model to enhance their arguments. 
However, what is worth mentioning is that some participants does not want examples to be shown at the beginning of their argument writing. The reason is that they thought the examples might distract their own opinions. We also observed that participants in Baseline focus more on analyzing the original post for wiring arguments to extract the key points or find the less rigorous opinions, which is also essential for persuasive arguments. Moreover, as suggested by some participants, unsuccessful arguments with low $\Delta$ points can also be presented as learning materials to learn what persuasive strategies are not applicable in the current topics. Therefore, we should provide users with more autonomy and control on how to use the examples, e.g., show or not show examples, show most relevant ones, or all examples.

\finalchange{\textbf{Reduce Cognitive Load by Prioritizing Semantic Relevant Examples}
Cognitive load is an essential factor for users deciding whether to use a system. We suggest ranking the examples based on the semantic similarity with the arguments the user wrote. The reason is that the content of the arguments is vital as the persuasive strategy. It took participants time to find suitable examples from the given strategy, even within the same topic. It will be easier to look for appropriate materials if semantic relevant examples are prioritized. In addition, a few participants (B5, B8) mentioned that when they selected one persuasive strategy, e.g., logos, in Example View, they found it tedious to scroll a lot to check the highlighted logos in all the examples. Therefore, we need to consider keeping the readability without losing the context of a paragraph by hiding all the other sentences or summarizing the context information, e.g., using keywords.}

\textbf{Provide visual representation and comparison of persuasive strategies for reflection.}
Our findings from user studies echo previous research that cognitive dissonance leads to reflection on people's thinking and behaviors~\cite{festinger1962cognitive, AL-CHI2020}. Instead of displaying one persuasiveness score as in ~\cite{AL-CHI2020}, we investigate the persuasiveness from the perspective of persuasive strategies users have applied. We also explicitly display the detailed differences in persuasive strategies compared to successful arguments for users to inspect. In particular, the left part of the chart(Figure~\ref{fig:usage1}(E3)) indicates that what is lacking in the user input regarding others' writing can arouse users' attention. We find that this chart led most participants in the Persua group to re-edit their arguments, as captured by the experiment video recording and interview feedback. However, unlike previous work~\cite{AL-CHI2020}, though helpful, we find that structural information, such as the visual presentation of the claim-premise relationship, is not enough in improving persuasiveness in writing online discussions. The arguments are usually short in the online discussion setting, with a few claims and several premises. The claim-premise relationship is relatively straightforward in this case, and the visual representation of the claim-premise structure provides limited information for users to reflect on and revise their arguments.

\subsection{Limitations and Future work}
Our work has several limitations that we can address in future work.

\xingbo{\textbf{Address the inaccuracies and uncertainties of AI algorithms.}
Although in the user study most participants found the algorithm's performance acceptable, some noticed some inaccuracies. Since this may discourage some users when errors happen~\cite{dietvorst2015algorithm}, we should further improve the current model in terms of accuracy or recall scores in argument components (Table~\ref{table:algorithmT}). We proposed the following potential solutions. First, label more data. We only labeled \sentence sentences.
Though lacking a valid dataset is a common problem in recent works~\cite{AL-CHI2020, wang2020argulens}, we need to enrich the dataset with more annotated texts. Besides, the current dataset is not balanced enough for each type of premise.
More sentences and cases need to be collected for a more balanced dataset. Second, more context information should be considered for the detection algorithm to classify sentences at a more flexible level rather than sentence-level. The reason is that one sentence sometimes have different meaning and indicate different strategies when combined with neighboring sentences. For example, ``\textit{It does not make sense.}'' can be \textit{logos} when it is in the middle of a paragraph of the logical reasoning and can also be \textit{pathos} when the context is trying to raise people's emotion of anger. Third, the algorithm may be hard to perfect. In the front-end, we can further notify users that the algorithm might not be 100\% all the time to convey the uncertainty or use transparency to encode the label accuracy. Besides, the system can provide a way to refine the results. The back-end algorithm can then be updated based on the labeled data, which is like the learner sourcing method introduced in ~\cite{kim2015learnersourcing}.}


\textbf{Adapt to the people being persuaded.}
Based on the user survey at the beginning of our study, participants mentioned that it is challenging to adapt their language style to the topic they are discussing to improve text persuasiveness. In the user study, we found that participants also wanted to know the personality and demographic information of the poster and try to adapt their language to the people they were trying to persuade. For example, in Task 1, the poster of the original post is a fifteen-year-old boy, and some participants were worried whether the poster could understand the arguments they wrote. Our system currently tries to solve the issue by providing examples from different well-accepted arguments while it is indirect and not always there are well-accepted arguments to learn. In the future, the intelligent system can allow users to customize the features of people they are going to persuade, and the corresponding language style and persuasive strategies should be generated and recommended for reference. 

\xingbo{\textbf{Generalize the system by considering long arguments and multi-turn arguments.}
We demonstrated the system's effectiveness in the scenario of online discussion forums. In fact, the design of illustrating examples of different persuasive strategies and comparing users' inputs with other arguments can be generalized to most of the other written contexts. Users could always be benefit from examples and different ways to compose persuasive strategies across various topics. However, different scenarios have their uniqueness. For instance,
in the essay writing context, arguments are much longer. Therefore, a more advanced algorithm could be developed to detect more complex claim-premise structures and relationships, e.g., the discourse path described in the work~\cite{liu2021exploring}. In the instant messaging scenario, arguments happen more dynamically with multiple rounds. Therefore, the detection of arguments mapping among multiple turns should be taken into consideration. For example, which claims or premises of the arguments from opponents have been addressed, how they are addressed, and when they are addressed in the dialogues are also exciting and promising areas to investigate. In addition, the way to learn persuasive strategies can also be more realistic with a dialog-based learning system as proposed in one recent work, Arguetutor~\cite{wambsganss2021arguetutor}.}

  
\section{Conclusion}
In this paper, we first derived four design goals of a tool that helps users improve the persuasiveness of arguments in online discussions through a survey with 123 online forum users and interviews with five debating experts. We then introduced, Persua, an interactive visualization system that assists people in enhancing the persuasiveness of their arguments in online discussion. 
Persua extends previous research by assisting adaptive arguments writing from the perspective of persuasive strategies. In particular, it provides good arguments (\emph{i.e.}, successfully persuade others in ChangeMyView\textsuperscript{\ref{CMV}}) for users to refer to; and compares these examples with users' own arguments in terms of persuasive strategies exploited. In addition, a between-subject user study with 36 participants shows that Persua can help people enhance persuasiveness in their arguments more compared to a baseline system. Finally, a set of design considerations have been summarized to guide the future design of the interactive systems that assist people in improving their persuasiveness in written texts.

\bibliographystyle{ACM-Reference-Format}
\bibliography{main}

\newpage
\xingbo{\appendix
\setcounter{table}{0}

\section{Model Selection for Argumentation Mining}
\label{appendix.model_selection}
After extracting argument-related features of training data using BERT, we input the features into several classic machine learning models and performed 5-fold stratified cross validation to evaluate and compare models on three tasks, including argument component extraction, argument relationship detection, and premise classification. 
\subsection{Argument Component Extraction}
\label{appendix.arg_compo}
The cross validation results of argument component extraction are summarized in Table~\ref{table.argu_component}.
The weighted average F1 score of Logistic Regression, Linear SVM, RBF SVM, Random Forest, Gaussian Naive Bayes, Nearest Neighbour, Adaboost Decision Tree are $0.78, 0.76, 0.77, 0.73, 0.70, 0.70, 0.74$, respectively. Thus, the final model we chose is Logistic Regression, which has the highest F1 score.

\begin{table}[b]
\caption{Evaluation of model performance of argument component extraction on training set using stratified 5-fold cross validation (* denotes the best model with the highest weighted average F1 score).}
\label{table.argu_component}
\begin{tabular}{lllll}
\hline
                                              &                   & \textbf{Precision} & \textbf{Recall} & \textbf{F1} \\ \hline
\multirow{3}{*}{\textbf{Logistic Regression *}} & Claim             & 0.67               & 0.25            & 0.37        \\
                                              & Premise           & 0.78               & 0.96            & 0.86        \\
                                              & Non-argumentative & 0.94               & 0.56            & 0.70        \\ \hline
\multirow{3}{*}{\textbf{Linear SVM}}          & Claim             & 0.73               & 0.15            & 0.24        \\
                                              & Premise           & 0.75               & 0.98            & 0.85        \\
                                              & Non-argumentative & 1.00               & 0.44            & 0.62        \\ \hline
\multirow{3}{*}{\textbf{RBF SVM}}             & Claim             & 0.83               & 0.18            & 0.30        \\
                                              & Premise           & 0.76               & 0.99            & 0.86        \\
                                              & Non-argumentative & 1.00               & 0.44            & 0.62        \\ \hline
\textbf{Random Forest}                        & Claim             & 0.17               & 0.02            & 0.03        \\
                                              & Premise           & 0.73               & 0.97            & 0.83        \\
                                              & Non-argumentative & 0.93               & 0.48            & 0.63        \\ \hline
\textbf{Gaussian NB}                          & Claim             & 0.40               & 0.29            & 0.34        \\
                                              & Premise           & 0.78               & 0.84            & 0.81        \\
                                              & Non-argumentative & 0.57               & 0.59            & 0.58        \\ \hline
\textbf{Nearest Neighbour}                    & Claim             & 0.40               & 0.29            & 0.34        \\
                                              & Premise           & 0.78               & 0.84            & 0.81        \\
                                              & Non-argumentative & 0.57               & 0.59            & 0.58        \\ \hline
\textbf{Adaboost Decision Tree}               & Claim             & 1.00               & 0.02            & 0.04        \\
                                              & Premise           & 0.73               & 0.99            & 0.84        \\
                                              & Non-argumentative & 0.81               & 0.48            & 0.60        \\ \hline
\end{tabular}
\end{table}

\subsection{Argument Relationship Detection}
\label{appendix.arg_relation}
The cross validation results of argument relationship detection are summarized in Table~\ref{table.argu_relation}.
The weighted average F1 score of Logistic Regression, Linear SVM, RBF SVM, Random Forest, Gaussian Naive Bayes, Nearest Neighbour, Adaboost Decision Tree are $0.92, 0.91, 0.91, 0.92, 0.85, 0.85, 0.91$, respectively. Thus, the final model we chose is Random Forest, which has the highest F1 score.

\begin{table}[]
\caption{Evaluation of model performance of argument relationship detection on training set using stratified 5-fold cross validation (* denotes the best model with the highest weighted average F1 score).}
\label{table.argu_relation}
\begin{tabular}{lllll}
\hline
                                                 &             & \textbf{Precision} & \textbf{Recall} & \textbf{F1} \\ \hline
\multirow{2}{*}{\textbf{Adaboost Decision Tree}} & Support     & 0.84               & 0.82            & 0.83        \\
                                                 & Non-support & 0.94               & 0.94            & 0.94        \\ \hline
\multirow{2}{*}{\textbf{Nearest Neighbour}}      & Support     & 0.87               & 0.52            & 0.65        \\
                                                 & Non-support & 0.85               & 0.97            & 0.91        \\ \hline
\multirow{2}{*}{\textbf{Gaussian NB}}            & Support     & 0.87               & 0.52            & 0.65        \\
                                                 & Non-support & 0.85               & 0.97            & 0.91        \\ \hline
\multirow{2}{*}{\textbf{Random Forest *}}        & Support     & 0.89               & 0.80            & 0.84        \\
                                                 & Non-support & 0.93               & 0.96            & 0.95        \\ \hline
\multirow{2}{*}{\textbf{Logistic Regression}}             & Support     & 0.87               & 0.80            & 0.83        \\
                                                 & Non-support & 0.93               & 0.96            & 0.94        \\ \hline
\multirow{2}{*}{\textbf{RBF SVM}}                & Support     & 0.85               & 0.80            & 0.82        \\
                                                 & Non-support & 0.93               & 0.95            & 0.94        \\ \hline
\multirow{2}{*}{\textbf{Linear SVM}}                      & Support     & 0.88               & 0.76            & 0.82        \\
                                                 & Non-support & 0.92               & 0.96            & 0.94        \\ \hline
\end{tabular}
\end{table}

\subsection{Premise Classification}
\label{appendix.arg_premise}
The cross validation results of premise classification are summarized in Table~\ref{table.premise_cls}.
The weighted average F1 score of Logistic Regression, Linear SVM, RBF SVM, Random Forest, Gaussian Naive Bayes, Nearest Neighbour, Adaboost Decision Tree are $0.70, 0.71, 0.71, 0.65, 0.67, 0.68, 0.68$, respectively. Thus, the final model we chose is Linear SVM, which has the highest F1 score. 

\begin{table}[]
\caption{Evaluation of model performance of premise classification on training set using stratified 5-fold cross validation (* denotes the best model with the highest weighted average F1 score).}
\label{table.premise_cls}
\begin{tabular}{lllll}
\hline
 &
  {\color[HTML]{333333} } &
  \multicolumn{1}{r}{{\color[HTML]{333333} \textbf{Precision}}} &
  \multicolumn{1}{r}{{\color[HTML]{333333} \textbf{Recall}}} &
  \multicolumn{1}{r}{{\color[HTML]{333333} \textbf{F1}}} \\ \hline
 &
  {\color[HTML]{333333} Logos} &
  {\color[HTML]{333333} 0.77} &
  {\color[HTML]{333333} 0.84} &
  {\color[HTML]{333333} 0.81} \\
 &
  {\color[HTML]{333333} Pathos} &
  {\color[HTML]{333333} 0.78} &
  {\color[HTML]{333333} 0.25} &
  {\color[HTML]{333333} 0.38} \\
 &
  {\color[HTML]{333333} Evidence} &
  {\color[HTML]{333333} 0.73} &
  {\color[HTML]{333333} 0.66} &
  {\color[HTML]{333333} 0.70} \\
\multirow{-4}{*}{\textbf{Linear SVM *}} &
  {\color[HTML]{333333} Ethos} &
  {\color[HTML]{333333} 1.00} &
  {\color[HTML]{333333} 0.60} &
  {\color[HTML]{333333} 0.75} \\ \hline
{\color[HTML]{333333} } &
  {\color[HTML]{333333} Logos} &
  {\color[HTML]{333333} 0.76} &
  {\color[HTML]{333333} 0.81} &
  {\color[HTML]{333333} 0.78} \\
{\color[HTML]{333333} } &
  {\color[HTML]{333333} Pathos} &
  {\color[HTML]{333333} 0.62} &
  {\color[HTML]{333333} 0.18} &
  {\color[HTML]{333333} 0.28} \\
{\color[HTML]{333333} } &
  {\color[HTML]{333333} Evidence} &
  {\color[HTML]{333333} 0.73} &
  {\color[HTML]{333333} 0.71} &
  {\color[HTML]{333333} 0.72} \\
\multirow{-4}{*}{{\color[HTML]{333333} \textbf{RBF SVM}}} &
  {\color[HTML]{333333} Ethos} &
  {\color[HTML]{333333} 1.00} &
  {\color[HTML]{333333} 0.55} &
  {\color[HTML]{333333} 0.71} \\ \hline
 &
  {\color[HTML]{333333} Logos} &
  {\color[HTML]{333333} 0.72} &
  {\color[HTML]{333333} 0.81} &
  {\color[HTML]{333333} 0.76} \\
 &
  {\color[HTML]{333333} Pathos} &
  {\color[HTML]{333333} 0.67} &
  {\color[HTML]{333333} 0.21} &
  {\color[HTML]{333333} 0.32} \\
 &
  {\color[HTML]{333333} Evidence} &
  {\color[HTML]{333333} 0.66} &
  {\color[HTML]{333333} 0.57} &
  {\color[HTML]{333333} 0.61} \\
\multirow{-4}{*}{\textbf{Logistic Regression}} &
  {\color[HTML]{333333} Ethos} &
  {\color[HTML]{333333} 1.00} &
  {\color[HTML]{333333} 0.60} &
  {\color[HTML]{333333} 0.75} \\ \hline
 &
  {\color[HTML]{333333} Logos} &
  {\color[HTML]{333333} 0.78} &
  {\color[HTML]{333333} 0.67} &
  {\color[HTML]{333333} 0.72} \\
 &
  {\color[HTML]{333333} Pathos} &
  {\color[HTML]{333333} 0.33} &
  {\color[HTML]{333333} 0.75} &
  {\color[HTML]{333333} 0.46} \\
 &
  {\color[HTML]{333333} Evidence} &
  {\color[HTML]{333333} 0.71} &
  {\color[HTML]{333333} 0.65} &
  {\color[HTML]{333333} 0.68} \\
\multirow{-4}{*}{\textbf{Random Forest}} &
  {\color[HTML]{333333} Ethos} &
  {\color[HTML]{333333} 0.50} &
  {\color[HTML]{333333} 0.90} &
  {\color[HTML]{333333} 0.64} \\ \hline
 &
  {\color[HTML]{333333} Logos} &
  {\color[HTML]{333333} 0.79} &
  {\color[HTML]{333333} 0.78} &
  {\color[HTML]{333333} 0.78} \\
 &
  {\color[HTML]{333333} Pathos} &
  {\color[HTML]{333333} 0.52} &
  {\color[HTML]{333333} 0.46} &
  {\color[HTML]{333333} 0.49} \\
 &
  {\color[HTML]{333333} Evidence} &
  {\color[HTML]{333333} 0.69} &
  {\color[HTML]{333333} 0.58} &
  {\color[HTML]{333333} 0.63} \\
\multirow{-4}{*}{\textbf{Gaussian NB}} &
  {\color[HTML]{333333} Ethos} &
  {\color[HTML]{333333} 0.90} &
  {\color[HTML]{333333} 0.45} &
  {\color[HTML]{333333} 0.60} \\ \hline
 &
  {\color[HTML]{333333} Logos} &
  {\color[HTML]{333333} 0.79} &
  {\color[HTML]{333333} 0.78} &
  {\color[HTML]{333333} 0.78} \\
 &
  {\color[HTML]{333333} Pathos} &
  {\color[HTML]{333333} 0.52} &
  {\color[HTML]{333333} 0.46} &
  {\color[HTML]{333333} 0.49} \\
 &
  {\color[HTML]{333333} Evidence} &
  {\color[HTML]{333333} 0.69} &
  {\color[HTML]{333333} 0.58} &
  {\color[HTML]{333333} 0.63} \\
\multirow{-4}{*}{\textbf{Nearest Neighbour}} &
  {\color[HTML]{333333} Ethos} &
  {\color[HTML]{333333} 0.90} &
  {\color[HTML]{333333} 0.45} &
  {\color[HTML]{333333} 0.60} \\ \hline
 &
  {\color[HTML]{333333} Logos} &
  0.74 &
  0.81 &
  0.78 \\
 &
  {\color[HTML]{333333} Pathos} &
  0.62 &
  0.29 &
  0.39 \\
 &
  {\color[HTML]{333333} Evidence} &
  0.69 &
  0.62 &
  0.66 \\
\multirow{-4}{*}{\textbf{Adaboost Decision Tree}} &
  {\color[HTML]{333333} Ethos} &
  0.92 &
  0.55 &
  0.69 \\ \hline
\end{tabular}
\end{table}
}
\end{document}